\documentclass[fleqn,usenatbib]{mnras}
 \pdfoutput=1

\usepackage[T1]{fontenc}
\usepackage{ae,aecompl}

\usepackage{graphicx}	
\usepackage{amsmath}	
\usepackage{ulem}       
\usepackage[dvipsnames]{xcolor}
\usepackage{newtxtext,newtxmath}

\newcommand{\um}{\mu\textnormal{m}}

\newcommand{\RE}{R_{\oplus}}
\newcommand{\ME}{M_{\oplus}}
\newcommand{\RS}{R_{\odot}}
\newcommand{\Psurf}{P_{\textnormal{surf}}}
\newcommand{\Pc}{P_{\textnormal{c}}}
\newcommand{\Tatm}{T_{\textnormal{atm}}}
\newcommand{\Rp}{R_{\textnormal{p}}}
\newcommand{\Rs}{R_{\star}}
\newcommand{\Mp}{M_{\textnormal{p}}}

\title[Biosignature detectability with JWST]{Detecting the proposed CH$_4$-CO$_2$ biosignature pair with the \textit{James Webb Space Telescope}: TRAPPIST-1e and the effect of cloud/haze}

\author[T.\ Mikal-Evans]{Thomas Mikal-Evans$^{1,2,\star}$ \\
  $^{1}$Max Planck Institute for Astronomy, K\"{o}nigstuhl 17, D-69117 Heidelberg, Germany\\
  $^{2}$MIT Kavli Institute for Astrophysics and Space Research, 77 Massachusetts Avenue, Cambridge, MA 02139, USA\\
  $^{\star}$tmevans@mpia.de\\  
}

\date{Accepted 2021 November 6. Received 2021 October 19; in original form 2021 July 10.}

\pubyear{2021}

\begin{document}
\label{firstpage}
\pagerange{\pageref{firstpage}--\pageref{lastpage}}
\maketitle

\begin{abstract}
It is widely anticipated that the James Webb Space Telescope (JWST) will be transformative for exoplanet studies. It has even been suggested that JWST could provide the first opportunity to search for biosignatures in an alien atmosphere using transmission spectroscopy. This claim is investigated, specifically for the proposed anoxic biosignature pair CH$_4$-CO$_2$. The most favourable known target is adopted (TRAPPIST-1e), with an assumed atmospheric composition similar to the Archean Earth. Compared to previous studies, a more systematic investigation of the effect that cloud/haze-layers have on the detectability of CH$_4$ and CO$_2$ is performed. In addition to a clear atmosphere scenario, cloud/haze-layers are considered at eight pressure levels between 600\,mbar and 1\,mbar. These pressures cover a plausible range for H$_2$O cloud and photochemical haze, based on observations of solar system atmospheres and physical models of tidally-locked planets such as TRAPPIST-1e, although no assumptions regarding the cloud/haze-layer composition are made in this study. For the clear atmosphere and cloud/haze-layer pressures of 600-100\,mbar, strong ($5\sigma$) detections of both CH$_4$ and CO$_2$ are found to be possible with approximately 5-10 co-added transits measured using the Near Infrared Spectrograph (NIRSpec) prism, assuming a dry stratosphere. However, approximately 30 co-added transits would be required to achieve the same result if a cloud/haze-layer is present at 10\,mbar. A cloud/haze-layer at 1\,mbar would prevent the detection of either molecule with the NIRSpec prism for observing programs up to 50 transits (>200 hours of JWST time), the maximum considered.
\end{abstract}


\section{Introduction} \label{sec:intro}

A number of gases or gas combinations have been proposed as atmospheric biosignatures \citep[for a review see][]{2018AsBio..18..663S}. For a biosphere similar to the modern Earth, this includes the simultaneous presence of O$_2$/O$_3$ and CH$_4$. Future space telescopes, such as \textit{LUVOIR} \citep{2018NatAs...2..605R} and the \textit{Origins Space Telescope} \citep{2018NatAs...2..596B}, will be designed with the explicit aim of detecting this biosignature pair in a sample of temperate rocky exoplanets. However, CH$_4$ may prove too challenging for a \textit{LUVOIR}-like telescope to detect for an atmosphere similar to that of the modern Earth, due to the weakness of predicted features across the 0.2-2.0$\um$ wavelength range \citep{2018JATIS...4c5001W,2019AJ....157..213K}. In the meantime, the \textit{James Webb Space Telescope} (JWST) could potentially detect CH$_4$ in the atmosphere of particularly favourable rocky exoplanets, but it most likely will not be able to detect O$_2$/O$_3$ \citep[e.g.][]{2019A&A...624A..49W}. Regarding the latter, \cite{2020NatAs...4..372F} have recently highlighted a previously overlooked O$_2$--X collision-induced-absorption band at $6.4\um$, which appears to be the most promising O$_2$ feature for detection with JWST. In practice, the same authors find that even for the especially favourable target TRAPPIST-1e, a $5\sigma$ detection of this spectral band would require co-adding $>700$ transits. To put such a program in context, each transit observation for TRAPPIST-1e would cost approximately 4.3 hours of JWST time (see below). More significantly, there will be fewer than 200 observable transits over the maximum JWST lifetime of 10 years, given the approximately 6 day orbital period of TRAPPIST-1e and the fact that the TRAPPIST-1 system is only accessible to JWST for approximately 100 days per year \citep{2020BAAS...52.0208G}. Likewise, the strong O$_3$ band at $\sim 10\,\um$ appears impractical to detect with JWST, even assuming a highly favourable target with an atmosphere similar to that of the modern Earth. For example, \cite{2021MNRAS.505.3562L} have estimated that for TRAPPIST-1e, 100 transit observations (430 hours) with JWST would be insufficient to detect the $10\,\um$ O$_3$ band at $3\sigma$ significance.

Biospheres differing from that of the modern Earth may produce distinctive gas combinations that are more readily detectable. Considering an anoxic atmosphere similar to that prevailing on the Archean Earth, \cite{2018SciA....4.5747K} proposed the simultaneous detection of CH$_4$ and CO$_2$ as constituting a biosignature, which would be strengthened if combined with a non-detection of CO. Subsequently, \cite{2018AJ....156..114K} (KT2018a) investigated the prospects for detecting the CH$_4$-CO$_2$ pair in the atmosphere of TRAPPIST-1e, assuming the same atmospheric composition as the Archean Earth. KT2018a concluded that this could be achieved by co-adding approximately 10 transit observations made with the JWST Near Infrared Spectrograph (NIRSpec) prism, which covers the $\sim$1-5$\,\um$ wavelength range. Furthermore, KT2018a reported that the result was not significantly affected by the presence cloud that truncated the transmission spectrum for pressures above 10\,mbar.

This paper presents an independent assessment of the prospects for detecting CH$_4$-CO$_2$ in the atmosphere of an Archean-like TRAPPIST-1e, inspired by the work of KT2018a. Of the currently known transiting rocky planets that may be habitable,\footnote{http://phl.upr.edu/projects/habitable-exoplanets-catalog} TRAPPIST-1 e and f have the highest expected signal-to-noise for transmission spectroscopy, according to the \textit{Exoplanet Atmosphere Observatory Table} maintained by the Space Telescope Science Institute.\footnote{https://catalogs.mast.stsci.edu/eaot} General circulation model (GCM) simulations also indicate that TRAPPIST-1e is more likely than TRAPPIST-1f to support liquid water on its surface \citep[e.g.][]{2017ApJ...839L...1W,2018A&A...612A..86T,2020GMD....13..707F,2020ApJ...894...84S}. For these reasons, it is reasonable to regard TRAPPIST-1e as the most favourable target available for conducting a biosignature search with JWST using transmission spectroscopy.

As elaborated on below, a point of difference setting this study apart from that of KT2018a is the methodology adopted for determining if a gas species is detected. Here, a Bayesian evidence framework is employed to quantify the confidence with which CH$_4$ and CO$_2$ could each be uniquely identified in the atmosphere of an Archean-like TRAPPIST-1e among a variety of plausible gas species that may be spectrally active. A second point of difference is that the present study investigates in greater detail how the presence of cloud and/or photochemical haze at the planetary day-night terminator might complicate detections. In addition, the present study adopts the latest system properties for TRAPPIST-1e from \cite{2021PSJ.....2....1A}.

\section{Methods} \label{sec:methods}

\begin{figure*}
\centering  
\includegraphics[width=\linewidth]{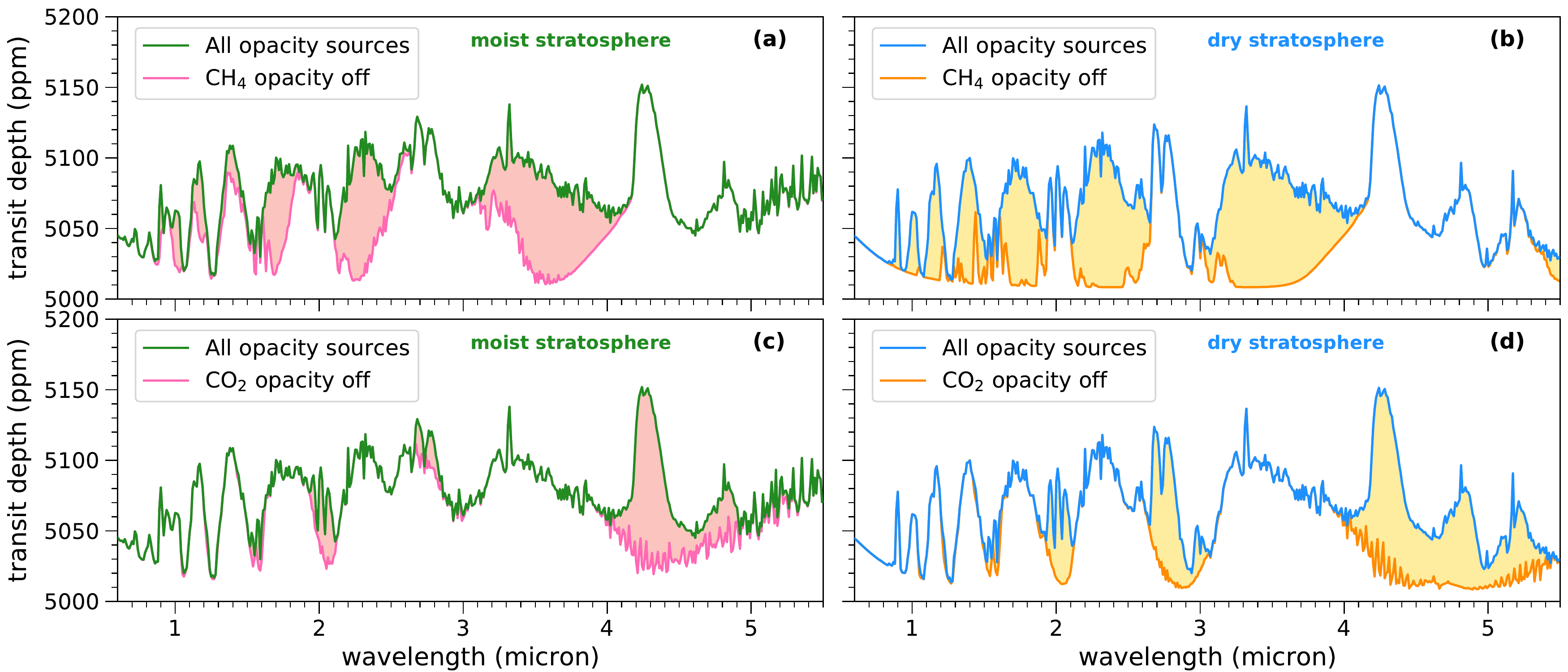}
\caption{Transmission spectra assuming a cloud-free atmosphere and Archean Earth composition. The moist atmosphere case is shown in panels (a) and (c) and the dry atmosphere case is shown in panels (b) and (d). For both cases, the top and bottom panels show, respectively, the same spectrum with the opacity of CH$_4$ and CO$_2$ switched off. Shading highlights the wavelengths at which each gas contributes uniquely to the absorption.}
\label{fig:absorbers}
\end{figure*}

To investigate the detectability of CH$_4$ and CO$_2$ in the atmosphere of TRAPPIST-1e assuming an Archean Earth composition, a hypothetical observing program with the NIRSpec prism is considered. The NIRSpec prism mode is chosen due to the broad wavelength coverage it affords ($0.8$-$5\um$), encompassing strong absorption features of both CH$_4$ and CO$_2$ as illustrated in Figure \ref{fig:absorbers}. Although it is conceivable that alternative instrument mode combinations may strictly be more efficient for detecting CH$_4$ and CO$_2$ (e.g.\ a hybrid NIRSpec program making use of both the prism and G395H grism), investigating such possibilities is beyond the scope of the present work. Furthermore, the broad wavelength coverage of the NIRSpec prism is likely to appeal to a wider range of scientific programs than the specific application considered here. The latter point is reinforced by the fact that of the approved Cycle 1 Guaranteed Time Observations (GTO) and General Observer (GO) programs that will be observing TRAPPIST-1 planets, those covering the 3-5\,$\um$ wavelength range will all be using the NIRSpec prism (GTO-1201, GTO-1331, GO-1981, GO-2420, GO-2589).

\subsection{Observation cost and synthetic datasets} \label{sec:datasets}

\begin{figure*}
\centering  
\includegraphics[width=\linewidth]{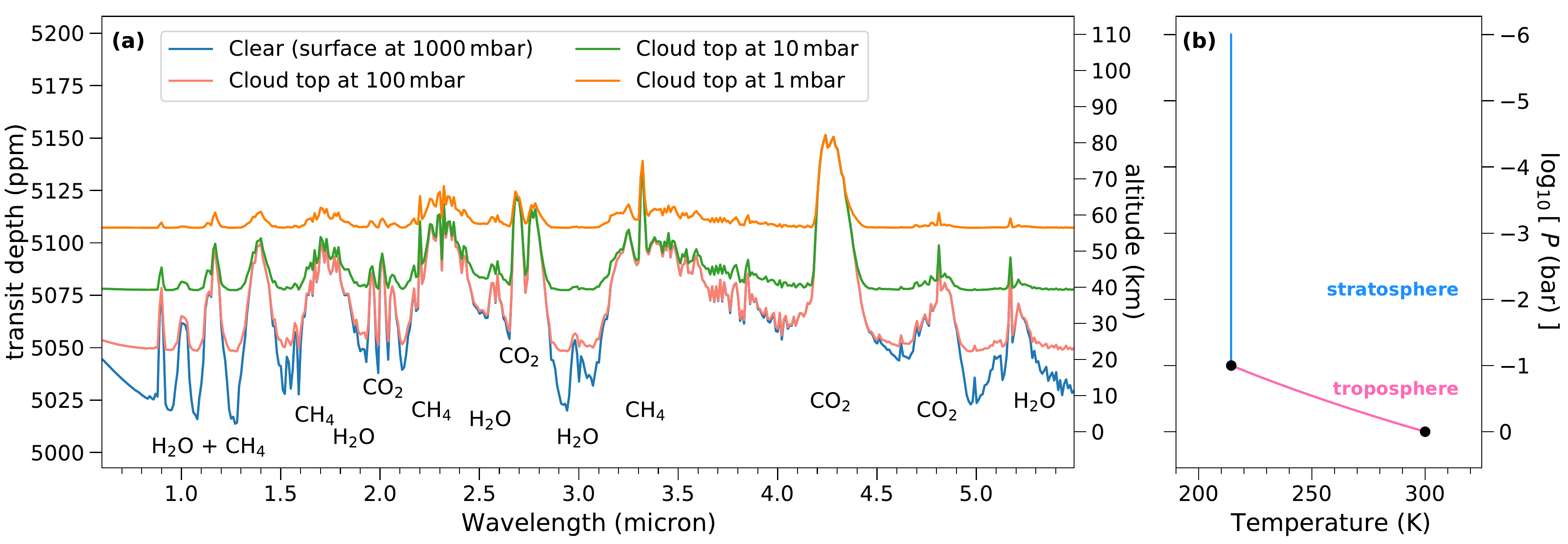}
\caption{(a) Moist atmosphere transmission spectra generated for different cloud-top assumptions with major absorption bands labelled. (b) The PT profile adopted in the atmospheric models that were used to generate all transmission spectra, with isothermal stratosphere and adiabatic troposphere.}
\label{fig:models}
\end{figure*}

The publicly available code \texttt{petitRADTRANS} \citep{2019A&A...627A..67M} was used to generate model transmission spectra for TRAPPIST-1e. An isothermal upper atmosphere with a temperature of $\Tatm=215$\,K was assumed for pressures below 100\,mbar, close to the mean stratosphere temperature of Earth. At higher pressures, a moist adiabatic temperature profile was adopted following \cite{2012ApJ...757..104R}, with the surface temperature set to 300\,K (Figure \ref{fig:models}). Two atmospheric compositions were considered, differing only in the assumed H$_2$O vapor abundance. The first `moist' scenario was identical to the Archean Earth composition used by KT2018a to enable direct comparison. For the spectrally active species, the $\log_{10}$ mole fractions were: $-2.301$\,dex for CH$_4$; $-1.301$\,dex for CO$_2$; $-8$\,dex for CO; and $-2$\,dex for H$_2$O. However, the transmission spectrum is primarily sensitive to gases in the stratosphere. For the present-day Earth atmosphere, most H$_2$O is cold-trapped below the tropopause, resulting in typical stratospheric H$_2$O mole fractions of a few ppm \citep[e.g.][]{doi:10.1029/97JD01371}. Similarly low stratospheric H$_2$O abundances likely prevailed during the Archean \citep[e.g.][]{2016AsBio..16..873A}. For this reason, a second `dry' scenario was considered, in which all abundances were the same as for the moist scenario, except the H$_2$O abundance was conservatively set to zero. Although the dry stratosphere scenario is more likely to coincide with habitable surface conditions, \cite{2017ApJ...845....5K} have identified cases in which habitable conditions are maintained with a moist stratosphere.

Under both the moist and dry scenarios, the remainder of the atmosphere was assumed to be composed of N$_2$, giving an atmosphere with mean molecular weight of approximately $29$ atomic mass units. The surface pressure was set to $\Psurf=1$\,bar and gases were assumed to be uniformly mixed throughout the atmosphere. A stellar radius of $\Rs=0.1192\,\RS$ was adopted \citep{2021PSJ.....2....1A}. For the planet, a radius of $\Rp=0.920\, \RE$ and mass of $\Mp=0.692\,\ME$ were used \citep{2021PSJ.....2....1A}, translating to a surface gravity of $8.0$\,m\,s$^{-2}$. Transmission spectra were generated for eight different assumptions of an opaque cloud-top pressure outlined below in Section \ref{sec:cloud}, as well as an idealised `clear' atmosphere without any cloud.

\subsubsection{Allowing partial saturation of the detector} \label{sec:psat}

The publicly available \texttt{pandexo} code \citep{2017PASP..129f4501B} was used to determine the noise properties associated with a the NIRSpec prism transit observation for TRAPPIST-1e, adopting the partial saturation strategy described in \cite{2018ApJ...856L..34B}. The resulting observing parameters for each transit were: SUB512 readout mode (i.e.\ $512 \times 32$ pixel subarray); NRSRAPID readout pattern; and 6 groups per integration with 6,515 integrations. This corresponds to a total observing time of 2.86\,h, which includes 0.932\,hr for the transit, 0.932\,hr for the minimal out-of-transit baseline, and an additional 1\,hr of out-of-transit baseline to avoid the 1\,hr penalty imposed for phase-constrained observations. Including observatory overheads, this translates to a total JWST cost of 4.3\,hr per TRAPPIST-1e transit, according to the Space Telscope Science Institute Astronomer's Proposal Tool (APT, version 2021.2).

\begin{figure}
\centering  
\includegraphics[width=\columnwidth]{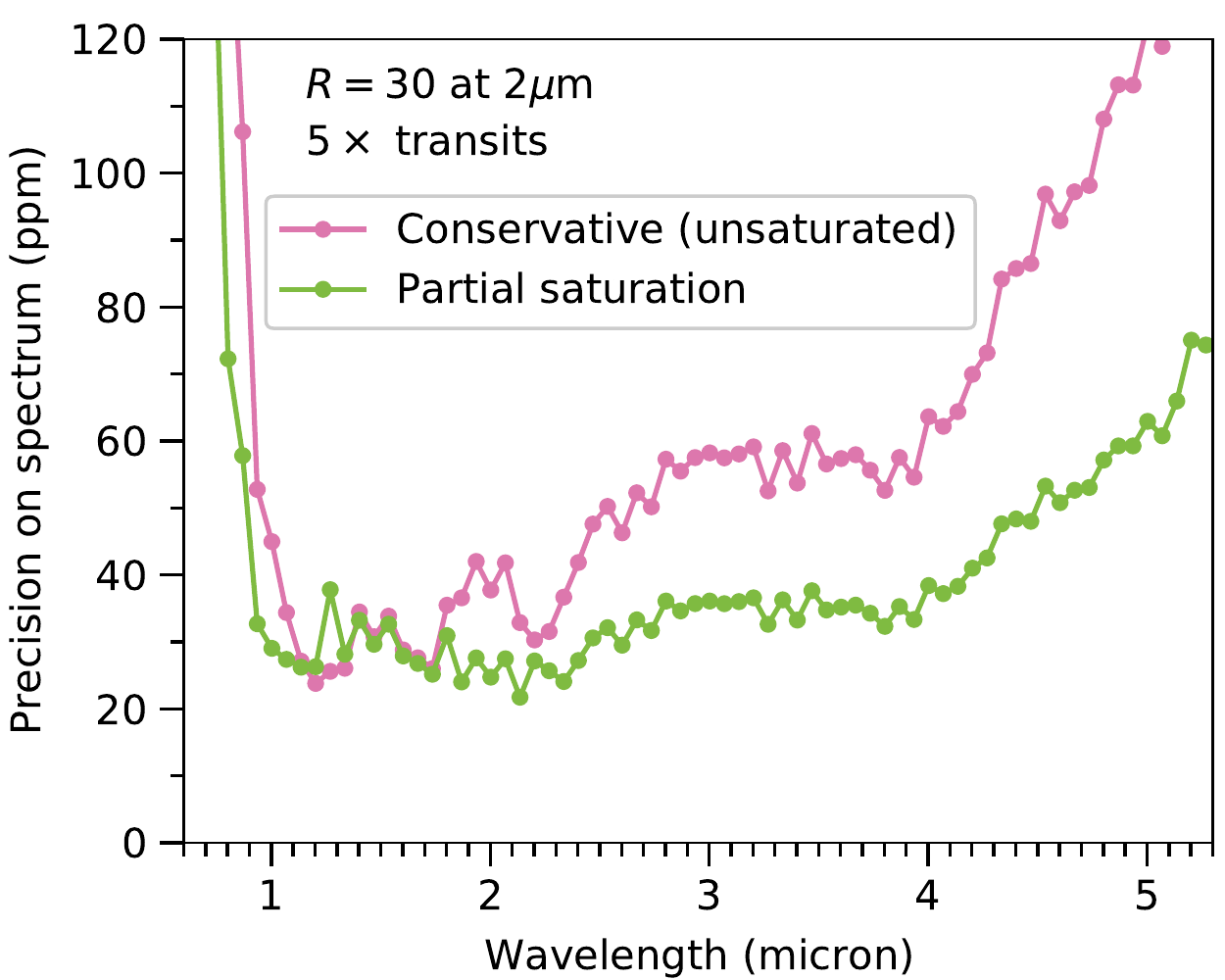}
\caption{Comparison of anticipated noise properties for the measured transmission spectrum of TRAPPIST-1e assuming a conservative detector read-out strategy that avoids saturation and a partial saturation read-out strategy. Similar to Figure 3 of \citet{2018ApJ...856L..34B}.}
\label{fig:noise}
\end{figure}

\begin{figure*}
\centering  
\includegraphics[width=\linewidth]{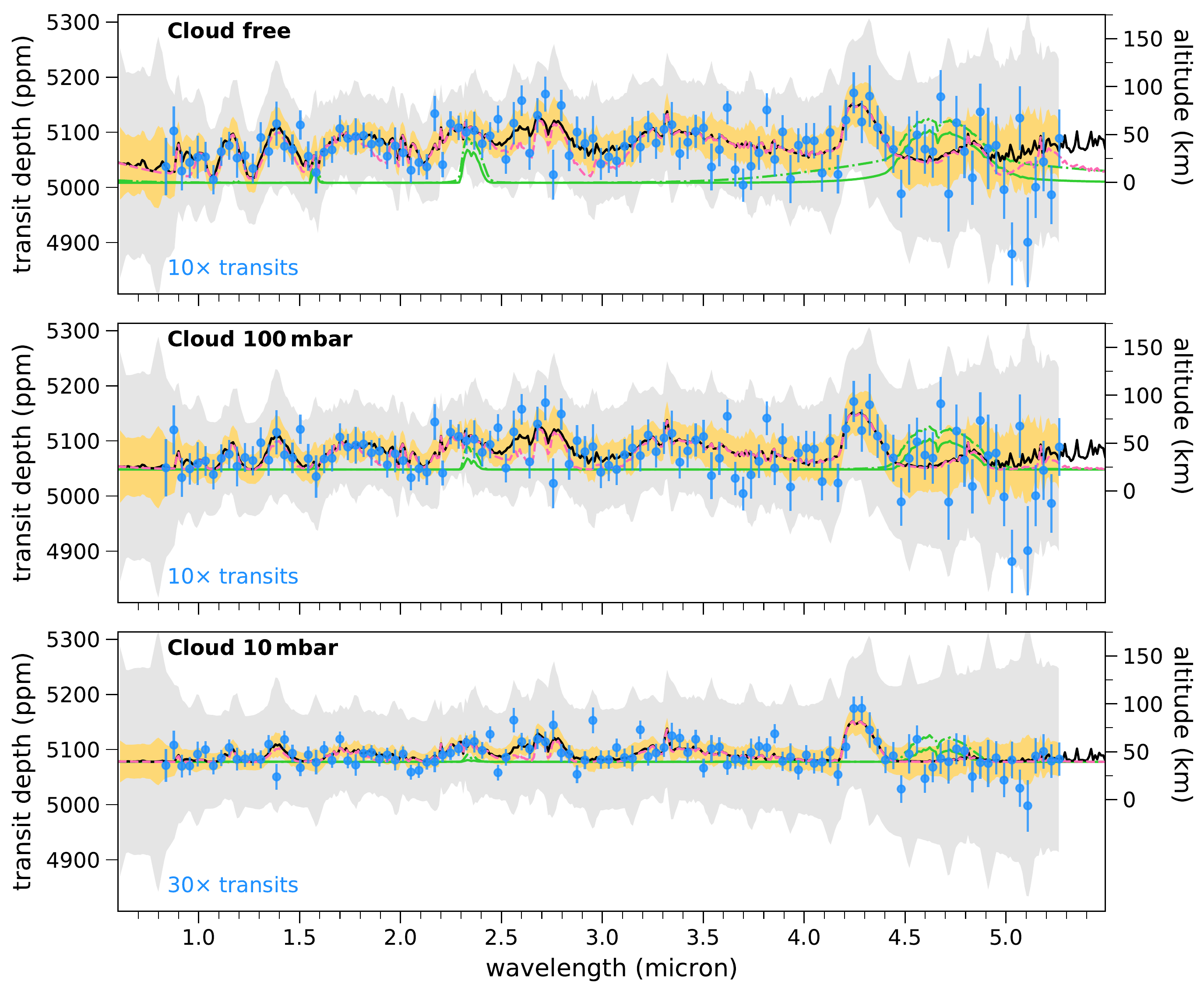}
\caption{Examples of synthetic datasets assuming a partial saturation observing strategy for different cloud-top scenarios assuming moist stratosphere models (solid black lines). Corresponding dry stratosphere models are also shown (dashed pink lines). The assumed number of transit observations is labelled in the lower left corner of each axis. Plotted datasets had the median $\chi^2$ from 1000 random noise realisations, representing a statistically typical outcome, and have been binned in wavelength to $R=100$ for visual clarity. In practice, retrieval analyses were performed on the models binned to the plotted spectral resolution ($R \sim 100$) without adding random noise perturbations, but still accounting for the measurement uncertainty as described in the text. Yellow shading indicates the measurement uncertainty at the resolution of the plotted data for the specified number of transits. Grey shading indicates the corresponding measurement uncertainties for a single transit observation.  Also shown are illustrative models with CO as the only active gaseous absorber at inflated mole fractions of -1.1\,dex (dot-dashed green lines) and -2.5\,dex (solid green lines), highlighting the difficulty of placing strong upper limits on the CO abundance using the NIRSpec prism in the case of an Archean-like atmosphere for TRAPPIST-1e.}
\label{fig:datasets}
\end{figure*}

Using the partial saturation strategy, 44 pixels were found to be saturated by the final (sixth) read of each integration. Noise properties for 5 and 39 pixels of the saturated pixels were calculated from the second and third non-destructive detector reads, respectively, by which point they had not saturated. This required applying a correction factor to account for the reduced duty cycle of the pixels saturating before the final read, as described in \cite{2018ApJ...856L..34B}. To illustrate the advantage of a partial saturation read-out strategy, Figure \ref{fig:noise} compares the predicted noise properties for the resulting transmission spectrum against those that would be obtained with a conservative read-out strategy that does not allow saturation. Example synthetic datasets generated using the \texttt{pandexo} noise properties are shown for the different cloud-top models in Figure \ref{fig:datasets}.

\subsection{Cloud-top altitudes}  \label{sec:cloud}

Evidence for condensible cloud and photochemically-produced haze has been uncovered in many of the exoplanet transmission spectra measured to date for gas giants and sub-Neptunes \citep[e.g.][]{2014Natur.505...69K,2016Natur.529...59S}. An optically thick layer of cloud or haze can substantially alter the detection prospects for absorption bands in the transmission spectrum, as can be appreciated by inspecting Figure \ref{fig:models}. Indeed, only a trace amount of cloud or haze is required to be a significant opacity source in a transmission spectrum due to the slant viewing geometry \citep{2005MNRAS.364..649F}. Note that unlike for an Earth-like planet transiting a Sun-like star \citep{2014ApJ...791....7B}, refraction has a negligible effect on the transmission spectra of planets orbiting within the much closer-in habitable zones of M dwarfs \citep{2017ApJ...850..128R}. As such, cloud/haze should be the primary source of atmospheric obscuration for TRAPPIST-1e.

For a temperate planet such as TRAPPIST-1e, the most plausible condensible species is H$_2$O. On average, approximately 70\% of the Earth is covered by H$_2$O cloud at any given time, with typical cloud-tops becoming optically thick at pressures in the $\sim 500$-$700$\,mbar range when observed from close to zenith \citep[e.g. see Figures 4 and 7 of ][]{EarthCloudKing2013}. This translates to cloud-top altitudes of approximately 3-6\,km.\footnote{U.S.\ Standard Atmosphere, 1976.} However, H$_2$O clouds could be mixed to higher altitudes in the atmospheres of close-in rocky planets such as TRAPPIST-1e, which have almost certainly been locked into synchronous orbits by tidal forces \citep[e.g.][]{2017CeMDA.129..509B}. Under this scenario, numerous 3D GCMs have found that typical cloud-top altitudes at the day-night terminator are likely to be around 15\,km, corresponding to pressures of approximately 100\,mbar \citep{2019ApJ...887..194F,2020ApJ...888L..20K,2020ApJ...891...58S,2020ApJ...898L..33P}. The only habitable zone exoplanet with a measured transmission spectrum precise enough to reveal spectral features -- K2-18b, a sub-Neptune orbiting an M dwarf -- appears to be cloudy, with a cloud-top pressure estimated between approximately $10$-$100$\,mbar, close to where the temperature profile crosses the H$_2$O condensation curve \citep{2019ApJ...887L..14B}. Although, it should be noted that the latter interpretation remains open to question, as \cite{2021arXiv210914608B} have recently argued that stellar inhomogeneities cannot yet be ruled out as an alternative explanation for the existing K2-18b data. For other transmission spectra published to date for planets in habitable zones, the data have only been precise enough to rule out cloud-free, H$_2$/He-dominated atmospheres \citep{2018NatAs...2..214D} or produced inconclusive results \citep{2021AJ....161...44E}. 

In addition to condensible clouds, an optically thick layer of hydrocarbon haze could form in the stratosphere of TRAPPIST-1e at pressures below 100\,mbar, driven by ultraviolet photochemistry. Hazes of this kind are thought to have been present in the atmosphere of Archean Earth \citep[e.g.][]{2016AsBio..16..873A}, given the favourably high abundance of CH$_4$, which could be broken up and recombined into longer hydrocarbon chains \citep{1985rapm.book...17A}. In the atmosphere of Titan, formation of an analogous photochemical haze is favoured by a reducing atmosphere with high CH$_4$ and N$_2$ abundances \citep{2017JGRE..122..432H}. Using a stellar occultation measurement, \cite{2014PNAS..111.9042R} demonstrated that the transmission spectrum for Titan's atmosphere is insensitive to pressures above approximately $0.1$-$10$\,mbar across the $0.9$-$5\um$ wavelength range due to this haze. 

For the present study, experiments were performed for eight different cloud-top pressures spanning the range of plausible scenarios outlined above: 600\,mbar (4\,km), 250\,mbar (11\,km), 120\,mbar (15\,km), 100\,mbar (16\,km), 50\,mbar (21\,km), 10\,mbar (31\,km), 5\,mbar (36\,km), and 1\,mbar (48\,km). In particular, the 600\,mbar case represents a conservative Earth-like H$_2$O cloud scenario. The 100\,mbar and 120\,mbar cases reflect more plausible H$_2$O cloud scenarios for a slowly-rotating planet such as TRAPPIST-1e. The 250\,mbar cloud/haze-layer is included as an intermediate case between the 600\,mbar and 100-120\,mbar physically-motivated scenarios, to provide additional coverage of the parameter space. The 10\,mbar case can be considered an optimistic Titan-like atmosphere, as it neglects the steep rise in opacity towards shorter wavelengths. Studying the atmosphere of TRAPPIST-1e specifically, \cite{2019ApJ...887..194F} modeled the production of photochemical haze assuming an Archean Earth composition and found it could become optically thick at effective altitudes spanning approximately 45-20\,km across the $1$-$5\um$ wavelength range (see their Figure 9). These altitudes correspond to pressures of approximately 1-100\,mbar, somewhat deeper than is observed for Titan. It is therefore also possible that an Archean haze would be intermediate between the 100\,mbar and 1\,mbar cloud-top cases considered here, although haze altitudes can be sensitive to numerous factors \citep{2019ApJ...877..109K}. In any case, the present study does not presume a specific origin for the 100\,mbar, 50\,mbar, 10\,mbar, 5\,mbar, and 1\,mbar cloud-tops, but presents them as representative cases covering a range of plausible scenarios. 

\subsection{Bayesian retrieval analyses} \label{sec:retrieval}

Ideally, an atmospheric biosignature detection would be made by convincingly demonstrating that a measured signal could only be plausibly explained by the specific gas/gases that constitute the biosignature. A distinction should be made between this fundamental first step and the subsequent process of constraining the abundances of the biosignature gases. In published studies investigating the detectability of gas species in temperate rocky exoplanets, it has been common practice to effectively skip the first step and assume the atmospheric composition is known \textit{a priori}. For example, a synthetic dataset is generated assuming some atmospheric composition, then a retrieval analysis or similar is performed with those same gases included and no others \citep[e.g.][]{2016MNRAS.461L..92B,2018ApJ...856L..34B,2018AJ....156..114K,2019AJ....158...27L}. In practice, however, it is preferable to remain agnostic about the atmospheric composition and avoid the possibility of a false positive, in which a combination of unrelated gases could conspire to produce a signal similar to that of the expected biosignature.

\begin{table*}
\begin{minipage}{\linewidth}
 \centering
 \caption{Median and $1\sigma$ credible ranges for CO$_2$ and CH$_4$ mole fractions, with corresponding $2\sigma$ and $3\sigma$ upper bounds for the CO mole fraction, assuming different cloud cases and number $N$ of co-added transits observed with the NIRSpec prism. \label{table:abundances}}

 \begin{tabular}{ccccccccccccc}
 \\ \hline
 \hline \multicolumn{2}{l}{Dry stratosphere} &&& & && \multicolumn{2}{l}{Moist stratosphere} &&& &  \\ \cline{1-2} \cline{8-9}
   &      &                && \multicolumn{2}{c}{CO upper (dex)} &&  &      &                && \multicolumn{2}{c}{CO upper (dex)}  \\ 

 Cloud       & $N$ &          CH$_4$ (dex) &         CO$_2$ (dex) & $2\sigma$ & $3\sigma$ &&  Cloud       & $N$ &         CH$_4$ (dex) &         CO$_2$ (dex) & $2\sigma$ & $3\sigma$   \\ \cline{1-6} \cline{8-13}
 \\
\smallskip Clear       & 10  & $-0.90_{-0.76}^{+0.52}$ & $-1.95_{-0.55}^{+0.47}$ & $<-2.53$ & $<-1.11$  &&  Clear      & 10  & $-1.08_{-0.88}^{+0.60}$ & $-2.05_{-0.60}^{+0.50}$ & $<-2.51$ & $<-1.27$  \\ 
\smallskip   600\,mbar & 10  & $-0.93_{-0.80}^{+0.54}$ & $-1.98_{-0.58}^{+0.50}$ & $<-2.47$ & $<-1.08$  &&  600\,mbar  & 10  & $-1.07_{-0.94}^{+0.61}$ & $-2.05_{-0.63}^{+0.51}$ & $<-2.63$ & $<-1.21$  \\ 
\smallskip   250\,mbar & 10  & $-0.93_{-0.80}^{+0.54}$ & $-1.94_{-0.62}^{+0.47}$ & $<-2.65$ & $<-1.06$  &&   250\,mbar & 10  & $-1.12_{-0.97}^{+0.67}$ & $-2.09_{-0.65}^{+0.54}$ & $<-2.68$ & $<-1.38$  \\ 
\smallskip   120\,mbar & 10  & $-0.97_{-0.91}^{+0.57}$ & $-1.99_{-0.65}^{+0.53}$ & $<-2.30$ & $<-0.93$  &&   120\,mbar & 10  & $-1.26_{-1.02}^{+0.73}$ & $-2.13_{-0.71}^{+0.60}$ & $<-2.94$ & $<-1.57$  \\ 
\smallskip   100\,mbar & 10  & $-0.96_{-0.92}^{+0.58}$ & $-1.97_{-0.67}^{+0.54}$ & $<-2.20$ & $<-0.64$  &&   100\,mbar & 10  & $-1.18_{-1.07}^{+0.70}$ & $-2.09_{-0.67}^{+0.55}$ & $<-2.53$ & $<-0.79$  \\
\smallskip   10\,mbar  & 10  & $-1.46_{-1.44}^{+0.88}$ & $-2.32_{-1.20}^{+0.80}$ & $<-1.12$ & $<-0.17$  &&   10\,mbar  & 10  & $-1.50_{-1.44}^{+0.89}$ & $-2.37_{-1.25}^{+0.78}$ & $<-1.22$ & $<-0.26$  \\
\smallskip             & 20  & $-1.16_{-0.99}^{+0.68}$ & $-2.16_{-0.68}^{+0.57}$ & $<-1.68$ & $<-0.33$  &&             & 20  & $-1.16_{-1.03}^{+0.68}$ & $-2.23_{-0.67}^{+0.54}$ & $<-1.99$ & $<-0.66$  \\
\smallskip             & 30  & $-1.09_{-0.97}^{+0.60}$ & $-2.16_{-0.63}^{+0.47}$ & $<-2.01$ & $<-0.70$  &&             & 30  & $-1.07_{-0.93}^{+0.61}$ & $-2.14_{-0.60}^{+0.43}$ & $<-2.17$ & $<-1.20$  \\ \hline
\end{tabular}                                                       
\end{minipage}
\end{table*}

To achieve this, the Bayesian framework described in \cite{2013ApJ...778..153B} is employed. This involves computing the Bayesian evidence for a `full' model including the suite of numerous possible gas species, then systematically computing the Bayesian evidence for models with one gas species removed at a time. By comparing the Bayesian evidences of the latter models to that of the full model, it is possible to compute the associated Bayes factors and thus determine the level of evidence for each individual gas species. Notably, \cite{2020ApJ...901L...1K} applied this approach in assessing the detectability of gas species in a temperate terrestrial planet orbiting a white dwarf. However, unlike the present study, those authors did not explicitly consider the effect of cloud on detectability, which as shown below, can be significant for transmission spectroscopy. \cite{2021MNRAS.505.3562L} also employed the same Bayesian approach to assess detectability of various molecules in the atmosphere of TRAPPIST-1e assuming prebiotic and modern Earth compositions, but only considered cloud-top altitudes of 6\,km and 12\,km. Another recent application was presented by \cite{2020AJ....159..117T} as part of a trade-space study for a hypothetical infrared observatory covering various wavelength ranges between 1-30$\,\um$. Those authors took TRAPPIST-1e as the target, but assumed a modern Earth-like atmospheric composition rather than the Archean-like composition considered here and only considered a single cloud-layer scenario with a cloud-top pressure of 560\,mbar.

For the full model considered in the present analysis, the spectrally active gas species H$_2$O, CH$_4$, CO, CO$_2$, NH$_3$, H$_2$S, HCN, and C$_2$H$_2$ were included as free parameters. The $\log_{10}$ mass fractions of these gases were allowed to vary uniformly between $[-10,0]$\,dex. For each model evaluation, the N$_2$ abundance was set to ensure the mass fractions summed to unity. The other free parameters were: the planetary mass ($M_p$) with Gaussian prior $M_p = 0.692 \pm 0.022 \, \ME$ \citep{2021PSJ.....2....1A}; a reference level planetary radius ($R_p$) with a uniform prior between $[0.8,1.1]$ as adopted by KT2018a and encompassing the posterior distribution $R_p = 0.920_{-0.012}^{+0.013} \, \RE$ reported by \cite{2021PSJ.....2....1A}; the $\log_{10}$ reference pressure ($\log_{10}P_0$) corresponding to the reference radius, with a uniform prior between $[-8,2]$\,dex bar; the $\log_{10}$ pressure of an opaque cloud deck ($\log_{10}\Pc$), also allowed to vary between $[-8,2]$\,dex bar; and the atmospheric temperature ($\Tatm$), assumed to be isothermal with a uniform prior between $[100,700]$ Kelvin. The stellar radius ($\Rs$) was held fixed to the value quoted in Section \ref{sec:datasets}.

The evidence for the full atmosphere model was also compared to that of a simple `flat' model, which assumed the planet lacks an atmosphere. For the flat model, the transmission spectrum is featureless, with the opaque planet radius ($R_p$) being the only free parameter. Although the full atmosphere model could also account for a featureless transmission spectrum by, for instance, placing a cloud deck at low pressures or decreasing abundances of spectrally active species to negligible levels, the flat model represents a physically plausible scenario with a smaller prior volume (i.e.\ fewer free parameters). In principle, the latter means that the flat model would be conservatively favoured by the Bayesian evidence, unless the data itself contains relatively strong evidence in favour of the full atmosphere model. Hence, a strong preference for the full model over a simple flat model is an additional prerequisite to be satisfied for claiming a biosignature detection.

To avoid sampling bias, random perturbations were not added to the synthetic datasets prior to performing the retrieval analyses, following \cite{2018AJ....155..200F}. Instead, the models described in Section \ref{sec:datasets} were simply binned to the data resolution and assumed to have the associated measurement uncertainties returned by \texttt{pandexo} (Figure \ref{fig:datasets}). Retrieval analyses were then performed using \texttt{petitRADTRANS} to generate model transmission spectra and \texttt{PyMultiNest} \citep{2014A&A...564A.125B} to evaluate the Bayesian evidences using nested sampling with 500 live points per retrieval. The `$n$-sigma' ($n\sigma$) detection significances for CH$_4$ and CO$_2$ were computed from the relevant Bayes factors, using the method outlined in \cite{2013ApJ...778..153B}. This process was repeated for different numbers of co-added transit observations between $N=5$ and $N=50$ in increments of 5, assuming the measurement uncertainties binned down optimally as $1/\sqrt{N}$.  However, observing programs of $N>50$ transits were not considered, primarily because this would equate to more than 200 hours of JWST time devoted to a single target using only one observing mode. Such a program would likely pose significant scheduling challenges, given that JWST can only observe the TRAPPIST-1 system for approximately 100 days per year \citep{2020BAAS...52.0208G}. The scheduling challenge could be especially pronounced if the community decides to undertake additional transit observations of TRAPPIST-1e with instrument modes beside the NIRSpec prism \citep[e.g.\ see][]{2020AJ....160...15G}. However, if $N>50$ transit observations with the NIRSpec prism are approved, the present analysis could be extended as appropriate in a future study.

\section{Results} \label{sec:results}

\begin{figure*}
\centering  
\includegraphics[width=0.78\linewidth]{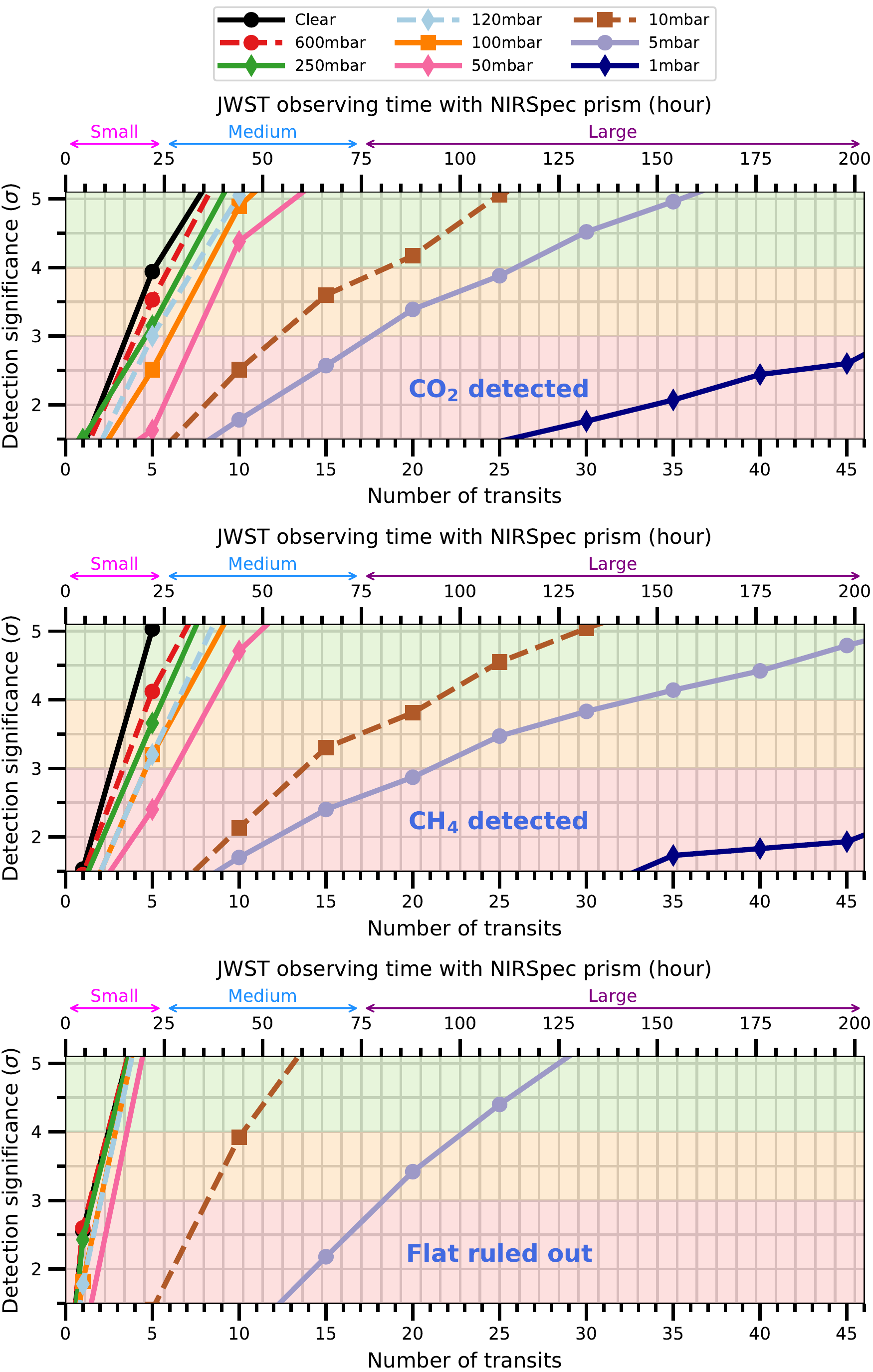}
\caption{Detection significance versus number of coadded transits for CO$_2$ (top panel) and CH$_4$ (middle panel) assuming a dry stratosphere. The bottom panel shows the significance with which the full atmosphere model is favoured over a flat transmission spectrum. For each panel, the top horizontal axis gives the JWST time required, assuming a time charge of 4.3\,hr per transit. Observing program categories of small ($<25$\,hr), medium (25-70\,hr), and Large ($>70$\,hr) are indicated. Note that the 1\,mbar cloud scenario falls below the displayed axis range in the bottom panel. Inspired by a similar figure in \citet{2020ApJ...901L...1K}. }
\label{fig:resultsDry}
\end{figure*}

\begin{figure*}
\centering  
\includegraphics[width=0.97\linewidth]{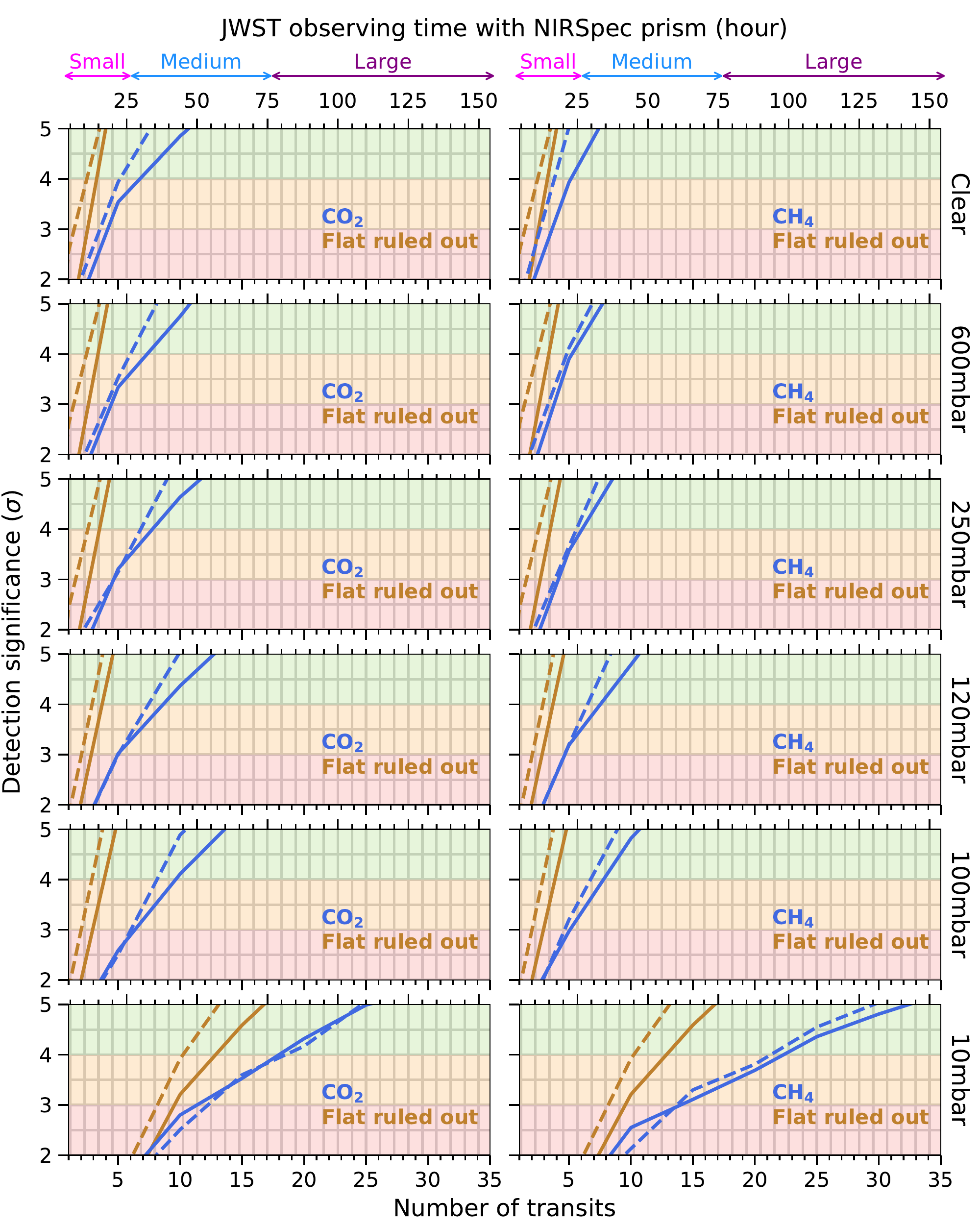}
\vspace{-10pt} 
\caption{Comparison of the detection significance versus observing time for the dry stratosphere (dashed lines) and moist stratosphere (solid lines) scenarios, with the same axis format as Figure \ref{fig:resultsDry}. Each row corresponds to a different cloud/haze assumption, as labelled along the right vertical axes. Blue lines indicate the detection significance of CO$_2$ (left column) and CH$_4$ (right column). A moist stratosphere generally increases the required observing time for CO$_2$ and CH$_4$ detections at a given significance, due to H$_2$O spectral bands overlapping those of CO$_2$ and CH$_4$. Brown lines indicate the significance with which the full atmosphere model is favoured over a flat transmission spectrum. The latter are identical for both columns of each row and are displayed to allow direct comparison with each molecule individually. }
\label{fig:resultsMoist}
\end{figure*}

The detection significances for each cloud case versus the number of co-added transits are shown in Figure \ref{fig:resultsDry}, assuming a dry stratosphere with the partial-saturation observing strategy described in Section \ref{sec:psat}. A direct comparison of the detection significances are shown for the dry and moist stratosphere scenarios in Figure \ref{fig:resultsMoist}. As expected, a moist stratosphere generally increases the required observing time to reach a given detection significance, due to H$_2$O spectral bands overlapping those of CH$_4$ and CO$_2$. However, the increased observing time typically corresponds to just 1-3 additional transits, depending on the molecule, cloud-top pressure, and detection significance (Figure \ref{fig:resultsMoist}).

Under the assumption of a dry stratosphere, strong ($5\sigma$) detections of both CH$_4$ and CO$_2$ are made for $N=5$-10 transits for cloud/haze-layer pressures down to 100\,mb (Figure \ref{fig:resultsDry}). For the specific case of a cloud-free atmosphere with moist stratosphere, $N \approx 10$ transits are found to be required for strong detections of both gases (Figure \ref{fig:resultsMoist}), in agreement with KT2018a. However, for the 10\,mbar cloud/haze-layer case, $N \approx 30$ transits are found to be necessary for ruling out the flat scenario described in Section \ref{sec:retrieval} and make $5\sigma$ detections of both CH$_4$ and CO$_2$ (Figures \ref{fig:resultsDry} and \ref{fig:resultsMoist}). This contrasts with KT2018a, who reported that the detectability of CH$_4$ and CO$_2$ was insensitive to the effect of a cloud layer if it were to truncate the transmission spectrum at the 10\,mbar pressure level. For the 5\,mbar cloud/haze-layer, identifying CH$_4$ becomes substantially more challenging, with $N \approx 35$ transits required for a $>4\sigma$ detection and $N>45$ transits required for a $5\sigma$ detection (Figure \ref{fig:resultsDry}). A cloud/haze-layer at 1\,mbar would likely prevent $>3\sigma$ detections of spectral features even with 200\,hr of JWST time, the maximum considered (Figure \ref{fig:resultsDry}). 

Posterior constraints for the retrieved mole fractions of CH$_4$, CO$_2$, and CO are reported in Table \ref{table:abundances} for the dry and moist stratosphere scenarios. Note that these mole fractions have been converted from the mass fractions that were used to compute the model transmission spectra during sampling (Section \ref{sec:retrieval}). In general, the CH$_4$ mole fraction constraints are $\sim 0.1$-0.2\,dex tighter for the dry stratosphere scenario compared to the moist stratosphere scenario, due to the overlap between CH$_4$ and H$_2$O spectral bands (Table \ref{table:abundances}). Since the strongest CO$_2$ spectral features are less degenerate with H$_2$O (Figure \ref{fig:models}), the CO$_2$ mole fraction constraints are essentially identical for the dry and moist stratosphere scenarios. Covariance plots for a subset of model parameters are shown in Figure \ref{fig:posterior2Dclear} for the clear atmosphere case with a dry stratosphere, assuming $N=10$ co-added transits. For all cases listed in Table \ref{table:abundances} except for the 10\,mbar cloud/haze-layer, the mole fractions of both CO$_2$ and CH$_4$ have $1\sigma$ uncertainties between $\sim 0.5$-1\,dex for $N=10$ co-added transits. For the 10\,mbar cloud/haze-layer, similar uncertainties are achieved for $N \geq 20$ co-added transits, and Figure \ref{fig:posterior2Dhz10mb} shows how the posterior constraints are affected by the assumption of $N=10$, $N=20$, and $N=30$ co-added transits.

\section{Discussion}

Assuming the atmosphere of TRAPPIST-1e has a composition similar to the Archean Earth, the results of this study indicate that strong ($5\sigma$) detections of both CH$_4$ and CO$_2$ could potentially be achieved by co-adding $N=5$-$10$ transit observations made with the NIRSpec prism (Figure \ref{fig:resultsDry}). This is in broad agreement with the results of KT2018a, who reported that $N \approx 10$ transits would be required to detect both molecules. However, unlike KT2018a, the presence of an optically thick cloud/haze layer at a pressure of $10$\,mbar is found to significantly increase the observing time required to $N \approx 30$ transits to achieve detections at the same significance level. A cloud/haze layer at 1\,mbar is found to prevent detections of both CH$_4$ and CO$_2$ even with 200\,hr observing time (Figures \ref{fig:resultsDry} and \ref{fig:resultsMoist}).

If CH$_4$ and CO$_2$ are detected together in the atmosphere of TRAPPIST-1e, KT2018a and KT2018b suggest that the interpretation of these gases as an Archean biosignature would be strengthened by constraining the CO mole fraction to below about $100\,$ppm. The present study finds that this likely cannot be achieved with the NIRSpec prism. Table \ref{table:abundances} reports $2\sigma$ and $3\sigma$ credible upper bounds for the CO mole fraction, which are at best around $-2.6$\,dex (2,500\,ppm) and $-1.6$\,dex (25,000\,ppm), respectively. The green lines in Figure \ref{fig:datasets} show the strength of CO absorption at these mole fractions, highlighting the challenge of using the NIRSpec prism to place strong upper limits on the CO mole fraction, at least for the case of an Archean-like atmosphere. However, complementary high-resolution spectroscopy measurements made with large ground-based telescopes may prove better suited for placing tighter upper limits on the CO mole fraction \citep[e.g.][]{2010Natur.465.1049S,2014A&A...565A.124B,2021Natur.592..205G}.

In another relevant study, \cite{2019ApJ...887..194F} (F2019) modeled the atmosphere of TRAPPIST-1e using a 3D general circulation model coupled to a 1D photochemistry model, assuming an Archean Earth composition. F2019 investigated the detectability of CH$_4$ and CO$_2$ using a relatively simple approach for evaluating detection significances, in which measurement uncertainties were directly compared to the amplitude of individual spectral features. Specifically, only the CH$_4$ band at $1.2\,\um$ and CO$_2$ band at $4.3\,\um$ were considered (Figure \ref{fig:absorbers}). This is different to the Bayesian methodology described in Section \ref{sec:retrieval}, which effectively accounts for all spectral features of a given species when considering if it can be detected. The updated \cite{2021PSJ.....2....1A} system parameters for TRAPPIST-1e adopted in the present study (Section \ref{sec:datasets}) were also not available at the time F2019 study. As with KT2018a, F2019 instead adopted the system parameters of \cite{2018A&A...613A..68G} (i.e. $\Mp=0.772\,\ME$, $\Rp=0.91 \RE$), corresponding to a surface gravity 14\% higher than assumed for the present study. Nonetheless, the results of F2019 appear to be broadly consistent with those of the present study. In particular, at wavelengths close to the $1.2\,\um$ CH$_4$ band, the F2019 model predicts an optically thick haze extending to altitudes of approximately $ 45$\,km, corresponding to pressures of $\sim 1$\,mbar. As for the 1\,mbar cloud/haze-layer case considered in the present study, F2019 find that CH$_4$ would be undetectable. In addition, the F2019 model predicts a decreasing haze opacity with increasing wavelength, lowering to an effective altitude of approximately $17$\,km ($\sim 100$\,mbar in pressure) at wavelengths close to the $4.3\,\um$ CO$_2$ band. F2019 find that a $3\sigma$ detection of CO$_2$ would require $N=8$ co-added transits, which is reasonably close to the $N\approx 6$ transits estimated by the present study (Figure \ref{fig:resultsMoist}). However, for a $5\sigma$ detection of CO$_2$, F2019 report that $N=23$ co-added transits would be required, which is significantly higher than the $N = 10$-$14$ co-added transits estimated by the present study, allowing for uncertainty in the moisture content of the stratosphere (Figure \ref{fig:resultsMoist}). This latter discrepancy could be due to a number of factors, such as the different methods used to evaluate detection significances, differences in the model transmission spectra (such as those arising due to the different assumed system properties), and F2019 perhaps not allowing for partial-saturation. To check the sensitivity of the results to these latter assumptions, additional calculations were performed using the \cite{2018A&A...613A..68G} parameters to generate the synthetic transmission spectra and without allowing for partial saturation (Figure \ref{fig:noise}). With these settings, the present study estimates that a $5\sigma$ detection of CO$_2$ for the 100\,mbar cloud/haze-layer would require $N \approx 35$ co-added transits, which is a factor of 2-3 higher than the fiducial estimate of $N=10$-$14$ co-added transits cited above, indicating that one or more such factors could potentially account for the different results obtained in this study and that of F2019. In any case, both F2019 and the present study conclude that cloud/haze at pressures below 100\,mbar would significantly hinder the detection CO$_2$ and CH$_4$.

\begin{figure}
\centering  
\includegraphics[width=\columnwidth]{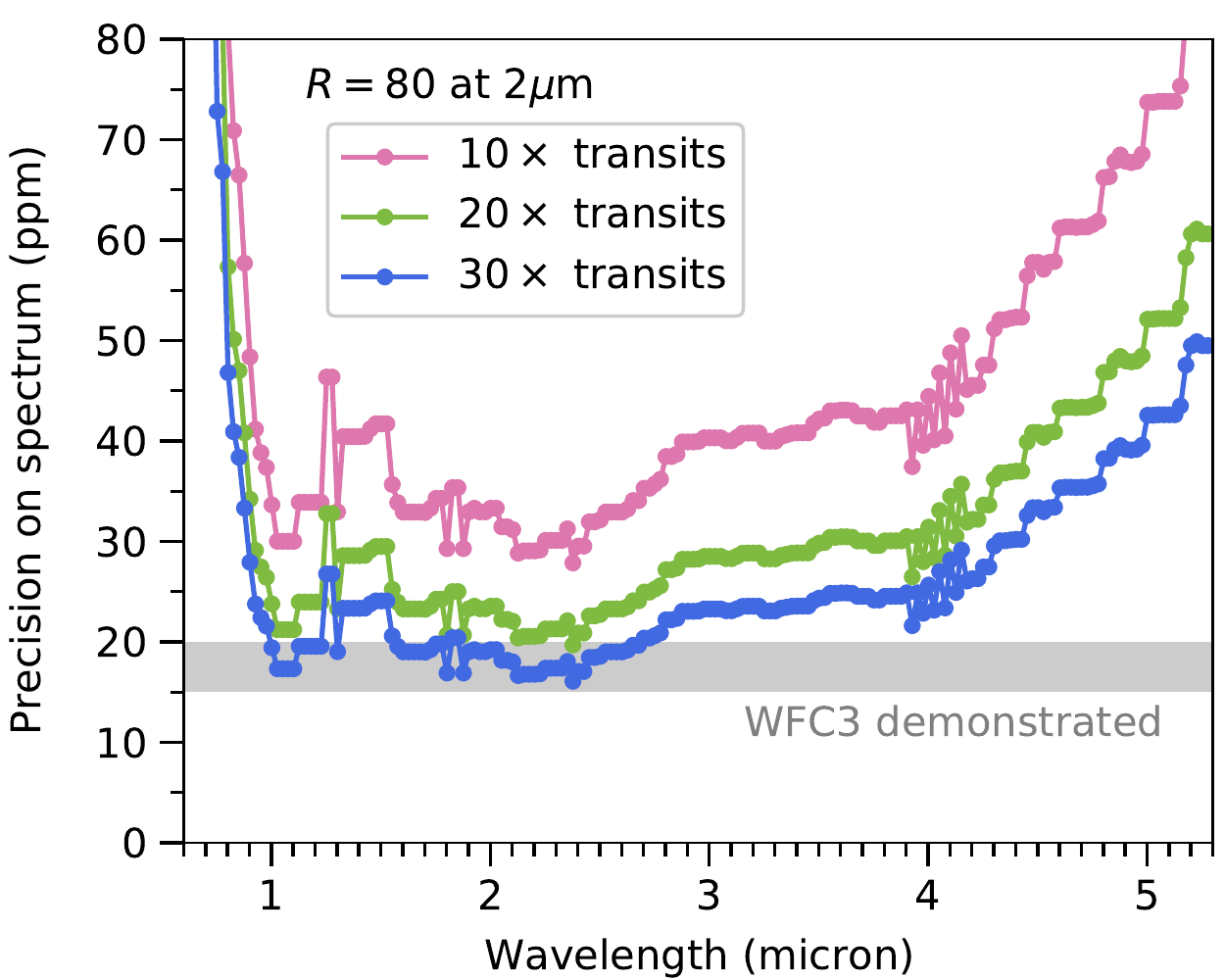}
\caption{Noise properties calculated with \texttt{pandexo} for the measured transmission spectrum of TRAPPIST-1e assuming a partial saturation observing strategy and different numbers of co-added transits. The grey region indicates precisions that have been demonstrated with HST WFC3 at the same spectral resolving power ($R=80$) by co-adding multiple, repeatable transit observations.}
\label{fig:noiseBin}
\end{figure}

Since $5\sigma$ detections of both CH$_4$ and CO$_2$ would required co-adding $N=30$ transits for the 10\,mbar cloud/haze-layer case (Figures \ref{fig:resultsDry} and \ref{fig:resultsMoist}), it is worth considering the corresponding measurement precisions that this would require of the NIRSpec prism. These precisions are shown in Figure \ref{fig:noiseBin} for $N=10$, $N=20$, and $N=30$ co-added transits, assuming the partial-saturation strategy described in Section \ref{sec:psat}. Also indicated is the measurement precision that has been demonstrated by co-adding transit observations made with the \textit{Hubble Space Telescope} (HST) Wide Field Camera 3 (WFC3) spectrograph. For example, measurement precisions of $\sim 15$-20\,ppm have been achieved across the 1.12-1.64\,um wavelength range with WFC3 at $R \sim 25$-$80$ by \cite{2016AJ....152..203L}, \cite{2021AJ....161...18M}, and \cite{2021AJ....161...19G}. In addition, these latter studies have demonstrated convincing transit-to-transit repeatability and resulting atmospheric spectra that are well explained by physically plausible models, which increases confidence in the quoted uncertainties being realistic. For $N=30$ co-added transits, Figure \ref{fig:noiseBin} shows that the highest precisions anticipated for NIRSpec will be comparable to these best precisions demonstrated with WFC3. It therefore seems reasonable to expect that NIRSpec could be capable of achieving these measurement precisions, given that NIRSpec will employ similar HgCdTe detectors to those of WFC3 \citep{2016ApJ...817...17G}. Indeed, the ultimate systematics noise floor of NIRSpec could even prove to be lower than that of WFC3, owing to the stable L2 orbit of JWST compared to the low-Earth orbit of HST. For example, the L2 orbit will allow continuous-stare observations to be made, which should help mitigate the charge-trapping systematic that affects WFC3 \citep{2017AJ....153..243Z}.  Ultimately, however, the on-sky performance of JWST is yet to be determined and the assumption of noise that bins down as $\sqrt{N}$ might prove too optimistic, somewhat analogously to early predictions of transit survey yields that neglected the role of systematic noise sources \citep[e.g.][]{2006MNRAS.373..231P}.

The results presented here also do not account for the possible effect of stellar variability. This could be especially relevant for the TRAPPIST-1 system, given that molecular absorption bands in star spots are possibly a significant source of contamination for the planetary transmission spectrum \citep[cf.\ \citealt{2019AJ....157...11W}]{2018AJ....156..178Z}. Another point to note is that each result shown in Figures \ref{fig:resultsDry} and \ref{fig:resultsMoist} effectively corresponds to the convergent mean of many repeated $N$-transit observing programs. In practice, only a single $N$-transit program would be conducted, representing a random draw from a distribution of $N$-transit programs. Exploring the dispersion of the latter would require repeating the present analysis many times, each with random measurement perturbations, which were not included here.

Finally, it should be stressed that many atmospheric compositions are possible for TRAPPIST-1e, a number of which have been investigated elsewhere \citep[e.g.][]{2016MNRAS.461L..92B,2017ApJ...850..121M,2019AJ....158...27L}. Only the very specific scenario of an Archean-like composition was considered here. However, the results obtained support the basic finding of KT2018a that, under this specific scenario, the detection of both CH$_4$ and CO$_2$ may be achievable with $N \leq 10$ transit observations made with the NIRSpec prism, corresponding to less than 50 hours of JWST time. Nonetheless, numerous uncertainties remain that cannot be resolved until such an observing program is executed. In particular, this includes the unknown presence or absence of a significant cloud/haze layer at pressures below 100\,mbar, which as noted above would either increase the amount of observing time required or prevent the detection of spectral features altogether.

\section{Conclusion}

If TRAPPIST-1e has an atmospheric composition similar to that of the Archean Earth, strong ($5\sigma$) detections for both CH$_4$ and CO$_2$ are possible for $N=5$-$10$ transit observations under the assumption of well-behaved instrumental noise and neglecting the effect of stellar variability. This result holds when cloud/haze-layers are included down to pressures of 100\,mbar (i.e.\ up to altitudes of $\sim 16$\,km), encompassing the range of H$_2$O cloud-top pressures predicted by GCM simulations of tidally-locked planets such as TRAPPIST-1e. However, if a cloud/haze-layer is present at a pressure level of 10\,mbar, the required observing time increases significantly to $N \approx 30$ transits for $5\sigma$ detections of both gases, assuming the NIRSpec prism will be able to achieve similar measurement precisions to the best demonstrated by HST WFC3. A cloud/haze-layer at the 1\,mbar pressure level would prevent the detection of both CH$_4$ and CO$_2$ even for a 200\,hr observing program devoted exclusively to TRAPPIST-1e.

\section*{Data Availability}

No new data were generated or analysed in support of this research.

\bibliographystyle{apj}
\bibliography{jwstbiosig}

\begin{thebibliography}{}
\expandafter\ifx\csname natexlab\endcsname\relax\def\natexlab#1{#1}\fi

\bibitem[{{Agol} {et~al.}(2021){Agol}, {Dorn}, {Grimm}, {Turbet}, {Ducrot},
  {Delrez}, {Gillon}, {Demory}, {Burdanov}, {Barkaoui}, {Benkhaldoun},
  {Bolmont}, {Burgasser}, {Carey}, {de Wit}, {Fabrycky}, {Foreman-Mackey},
  {Haldemann}, {Hernandez}, {Ingalls}, {Jehin}, {Langford}, {Leconte},
  {Lederer}, {Luger}, {Malhotra}, {Meadows}, {Morris}, {Pozuelos}, {Queloz},
  {Raymond}, {Selsis}, {Sestovic}, {Triaud}, \& {Van
  Grootel}}]{2021PSJ.....2....1A}
{Agol}, E., {Dorn}, C., {Grimm}, S.~L., {et~al.} 2021, {Planet. Sci. J.}, 2, 1

\bibitem[{{Arney} {et~al.}(2016){Arney}, {Domagal-Goldman}, {Meadows}, {Wolf},
  {Schwieterman}, {Charnay}, {Claire}, {H{\'e}brard}, \&
  {Trainer}}]{2016AsBio..16..873A}
{Arney}, G., {Domagal-Goldman}, S.~D., {Meadows}, V.~S., {et~al.} 2016,
  Astrobiology, 16, 873

\bibitem[{{Atreya} \& {Romani}(1985)}]{1985rapm.book...17A}
{Atreya}, S.~K., \& {Romani}, P.~N. 1985, {Photochemistry and clouds of
  Jupiter, Saturn and Uranus.}, ed. G.~E. {Hunt}, 17--68

\bibitem[{{Barclay} {et~al.}(2021){Barclay}, {Kostov}, {Col{\'o}n}, {Quintana},
  {Schlieder}, {Louie}, {Gilbert}, \& {Mullally}}]{2021arXiv210914608B}
{Barclay}, T., {Kostov}, V.~B., {Col{\'o}n}, K.~D., {et~al.} 2021, arXiv
  e-prints, arXiv:2109.14608

\bibitem[{{Barnes}(2017)}]{2017CeMDA.129..509B}
{Barnes}, R. 2017, Celestial Mechanics and Dynamical Astronomy, 129, 509

\bibitem[{{Barstow} \& {Irwin}(2016)}]{2016MNRAS.461L..92B}
{Barstow}, J.~K., \& {Irwin}, P.~G.~J. 2016, \mnras, 461, L92

\bibitem[{{Batalha} {et~al.}(2018){Batalha}, {Lewis}, {Line}, {Valenti}, \&
  {Stevenson}}]{2018ApJ...856L..34B}
{Batalha}, N.~E., {Lewis}, N.~K., {Line}, M.~R., {Valenti}, J., \& {Stevenson},
  K. 2018, \apjl, 856, L34

\bibitem[{{Batalha} {et~al.}(2017){Batalha}, {Mandell}, {Pontoppidan},
  {Stevenson}, {Lewis}, {Kalirai}, {Earl}, {Greene}, {Albert}, \&
  {Nielsen}}]{2017PASP..129f4501B}
{Batalha}, N.~E., {Mandell}, A., {Pontoppidan}, K., {et~al.} 2017, \pasp, 129,
  064501

\bibitem[{{Battersby} {et~al.}(2018){Battersby}, {Armus}, {Bergin}, {Kataria},
  {Meixner}, {Pope}, {Stevenson}, {Cooray}, {Leisawitz}, {Scott}, {Bauer},
  {Bradford}, {Ennico}, {Fortney}, {Kaltenegger}, {Melnick}, {Milam},
  {Narayanan}, {Padgett}, {Pontoppidan}, {Roellig}, {Sandstrom}, {Su},
  {Vieira}, {Wright}, {Zmuidzinas}, {Staguhn}, {Sheth}, {Benford}, {Mamajek},
  {Neff}, {Carey}, {Burgarella}, {De Beck}, {Gerin}, {Helmich}, {Moseley},
  {Sakon}, \& {Wiedner}}]{2018NatAs...2..596B}
{Battersby}, C., {Armus}, L., {Bergin}, E., {et~al.} 2018, Nature Astronomy, 2,
  596

\bibitem[{{Benneke} \& {Seager}(2013)}]{2013ApJ...778..153B}
{Benneke}, B., \& {Seager}, S. 2013, \apj, 778, 153

\bibitem[{{Benneke} {et~al.}(2019){Benneke}, {Wong}, {Piaulet}, {Knutson},
  {Lothringer}, {Morley}, {Crossfield}, {Gao}, {Greene}, {Dressing},
  {Dragomir}, {Howard}, {McCullough}, {Kempton}, {Fortney}, \&
  {Fraine}}]{2019ApJ...887L..14B}
{Benneke}, B., {Wong}, I., {Piaulet}, C., {et~al.} 2019, \apjl, 887, L14

\bibitem[{{B{\'e}tr{\'e}mieux} \& {Kaltenegger}(2014)}]{2014ApJ...791....7B}
{B{\'e}tr{\'e}mieux}, Y., \& {Kaltenegger}, L. 2014, \apj, 791, 7

\bibitem[{{Brogi} {et~al.}(2014){Brogi}, {de Kok}, {Birkby}, {Schwarz}, \&
  {Snellen}}]{2014A&A...565A.124B}
{Brogi}, M., {de Kok}, R.~J., {Birkby}, J.~L., {Schwarz}, H., \& {Snellen},
  I.~A.~G. 2014, \aap, 565, A124

\bibitem[{{Buchner} {et~al.}(2014){Buchner}, {Georgakakis}, {Nandra}, {Hsu},
  {Rangel}, {Brightman}, {Merloni}, {Salvato}, {Donley}, \&
  {Kocevski}}]{2014A&A...564A.125B}
{Buchner}, J., {Georgakakis}, A., {Nandra}, K., {et~al.} 2014, \aap, 564, A125

\bibitem[{Chiou {et~al.}(1997)Chiou, McCormick, \& Chu}]{doi:10.1029/97JD01371}
Chiou, E.~W., McCormick, M.~P., \& Chu, W.~P. 1997, Journal of Geophysical
  Research: Atmospheres, 102, 19105

\bibitem[{{de Wit} {et~al.}(2018){de Wit}, {Wakeford}, {Lewis}, {Delrez},
  {Gillon}, {Selsis}, {Leconte}, {Demory}, {Bolmont}, {Bourrier}, {Burgasser},
  {Grimm}, {Jehin}, {Lederer}, {Owen}, {Stamenkovi{\'c}}, \&
  {Triaud}}]{2018NatAs...2..214D}
{de Wit}, J., {Wakeford}, H.~R., {Lewis}, N.~K., {et~al.} 2018, Nature
  Astronomy, 2, 214

\bibitem[{{Edwards} {et~al.}(2021){Edwards}, {Changeat}, {Mori}, {Anisman},
  {Morvan}, {Yip}, {Tsiaras}, {Al-Refaie}, {Waldmann}, \&
  {Tinetti}}]{2021AJ....161...44E}
{Edwards}, B., {Changeat}, Q., {Mori}, M., {et~al.} 2021, \aj, 161, 44

\bibitem[{{Fauchez} {et~al.}(2019){Fauchez}, {Turbet}, {Villanueva}, {Wolf},
  {Arney}, {Kopparapu}, {Lincowski}, {Mandell}, {de Wit}, {Pidhorodetska},
  {Domagal-Goldman}, \& {Stevenson}}]{2019ApJ...887..194F}
{Fauchez}, T.~J., {Turbet}, M., {Villanueva}, G.~L., {et~al.} 2019, \apj, 887,
  194

\bibitem[{{Fauchez} {et~al.}(2020{\natexlab{a}}){Fauchez}, {Villanueva},
  {Schwieterman}, {Turbet}, {Arney}, {Pidhorodetska}, {Kopparapu}, {Mandell},
  \& {Domagal-Goldman}}]{2020NatAs...4..372F}
{Fauchez}, T.~J., {Villanueva}, G.~L., {Schwieterman}, E.~W., {et~al.}
  2020{\natexlab{a}}, Nature Astronomy, 4, 372

\bibitem[{{Fauchez} {et~al.}(2020{\natexlab{b}}){Fauchez}, {Turbet}, {Wolf},
  {Boutle}, {Way}, {Del Genio}, {Mayne}, {Tsigaridis}, {Kopparapu}, {Yang},
  {Forget}, {Mand ell}, \& {Domagal Goldman}}]{2020GMD....13..707F}
{Fauchez}, T.~J., {Turbet}, M., {Wolf}, E.~T., {et~al.} 2020{\natexlab{b}},
  Geoscientific Model Development, 13, 707

\bibitem[{{Feng} {et~al.}(2018){Feng}, {Robinson}, {Fortney}, {Lupu}, {Marley},
  {Lewis}, {Macintosh}, \& {Line}}]{2018AJ....155..200F}
{Feng}, Y.~K., {Robinson}, T.~D., {Fortney}, J.~J., {et~al.} 2018, \aj, 155,
  200

\bibitem[{{Fortney}(2005)}]{2005MNRAS.364..649F}
{Fortney}, J.~J. 2005, \mnras, 364, 649

\bibitem[{{Giacobbe} {et~al.}(2021){Giacobbe}, {Brogi}, {Gandhi}, {Cubillos},
  {Bonomo}, {Sozzetti}, {Fossati}, {Guilluy}, {Carleo}, {Rainer},
  {Harutyunyan}, {Borsa}, {Pino}, {Nascimbeni}, {Benatti}, {Biazzo},
  {Bignamini}, {Chubb}, {Claudi}, {Cosentino}, {Covino}, {Damasso}, {Desidera},
  {Fiorenzano}, {Ghedina}, {Lanza}, {Leto}, {Maggio}, {Malavolta}, {Maldonado},
  {Micela}, {Molinari}, {Pagano}, {Pedani}, {Piotto}, {Poretti}, {Scandariato},
  {Yurchenko}, {Fantinel}, {Galli}, {Lodi}, {Sanna}, \&
  {Tozzi}}]{2021Natur.592..205G}
{Giacobbe}, P., {Brogi}, M., {Gandhi}, S., {et~al.} 2021, \nat, 592, 205

\bibitem[{{Gillon} {et~al.}(2020){Gillon}, {Meadows}, {Agol}, {Burgasser},
  {Deming}, {Doyon}, {Fortney}, {Kreidberg}, {Owen}, {Selsis}, {de Wit},
  {Lustig-Yaeger}, \& {Rackham}}]{2020BAAS...52.0208G}
{Gillon}, M., {Meadows}, V., {Agol}, E., {et~al.} 2020, in Bulletin of the
  American Astronomical Society, Vol.~52, 0208

\bibitem[{{Greene} {et~al.}(2016){Greene}, {Line}, {Montero}, {Fortney},
  {Lustig-Yaeger}, \& {Luther}}]{2016ApJ...817...17G}
{Greene}, T.~P., {Line}, M.~R., {Montero}, C., {et~al.} 2016, \apj, 817, 17

\bibitem[{{Grimm} {et~al.}(2018){Grimm}, {Demory}, {Gillon}, {Dorn}, {Agol},
  {Burdanov}, {Delrez}, {Sestovic}, {Triaud}, {Turbet}, {Bolmont}, {Caldas},
  {de Wit}, {Jehin}, {Leconte}, {Raymond}, {Van Grootel}, {Burgasser}, {Carey},
  {Fabrycky}, {Heng}, {Hernandez}, {Ingalls}, {Lederer}, {Selsis}, \&
  {Queloz}}]{2018A&A...613A..68G}
{Grimm}, S.~L., {Demory}, B.-O., {Gillon}, M., {et~al.} 2018, \aap, 613, A68

\bibitem[{{Guilluy} {et~al.}(2021){Guilluy}, {Gressier}, {Wright}, {Santerne},
  {Jaziri}, {Edwards}, {Changeat}, {Modirrousta-Galian}, {Skaf}, {Al-Refaie},
  {Baeyens}, {Bieger}, {Blain}, {Kiefer}, {Morvan}, {Mugnai}, {Pluriel},
  {Poveda}, {Zingales}, {Whiteford}, {Yip}, {Charnay}, {Leconte}, {Drossart},
  {Sozzetti}, {Marcq}, {Tsiaras}, {Venot}, {Waldmann}, \&
  {Beaulieu}}]{2021AJ....161...19G}
{Guilluy}, G., {Gressier}, A., {Wright}, S., {et~al.} 2021, \aj, 161, 19

\bibitem[{{Guzm{\'a}n-Mesa} {et~al.}(2020){Guzm{\'a}n-Mesa}, {Kitzmann},
  {Fisher}, {Burgasser}, {Hoeijmakers}, {M{\'a}rquez-Neila}, {Grimm},
  {Mandell}, {Sznitman}, \& {Heng}}]{2020AJ....160...15G}
{Guzm{\'a}n-Mesa}, A., {Kitzmann}, D., {Fisher}, C., {et~al.} 2020, \aj, 160,
  15

\bibitem[{{Hinton}(2016)}]{2016JOSS....1...45H}
{Hinton}, S.~R. 2016, The Journal of Open Source Software, 1, 00045

\bibitem[{{H{\"o}rst}(2017)}]{2017JGRE..122..432H}
{H{\"o}rst}, S.~M. 2017, Journal of Geophysical Research (Planets), 122, 432

\bibitem[{{Kaltenegger} {et~al.}(2020){Kaltenegger}, {MacDonald}, {Kozakis},
  {Lewis}, {Mamajek}, {McDowell}, \& {Vanderburg}}]{2020ApJ...901L...1K}
{Kaltenegger}, L., {MacDonald}, R.~J., {Kozakis}, T., {et~al.} 2020, \apjl,
  901, L1

\bibitem[{{Kawashima} \& {Ikoma}(2019)}]{2019ApJ...877..109K}
{Kawashima}, Y., \& {Ikoma}, M. 2019, \apj, 877, 109

\bibitem[{{Kawashima} \& {Rugheimer}(2019)}]{2019AJ....157..213K}
{Kawashima}, Y., \& {Rugheimer}, S. 2019, \aj, 157, 213

\bibitem[{{King} {et~al.}(2013){King}, {Platnick}, {Menzel}, {Ackerman}, \&
  {Hubanks}}]{EarthCloudKing2013}
{King}, M.~D., {Platnick}, S., {Menzel}, W.~P., {Ackerman}, S.~A., \&
  {Hubanks}, P.~A. 2013, IEEE Transactions on Geoscience and Remote Sensing,
  51, 3826

\bibitem[{{Komacek} {et~al.}(2020){Komacek}, {Fauchez}, {Wolf}, \&
  {Abbot}}]{2020ApJ...888L..20K}
{Komacek}, T.~D., {Fauchez}, T.~J., {Wolf}, E.~T., \& {Abbot}, D.~S. 2020,
  \apjl, 888, L20

\bibitem[{{Kopparapu} {et~al.}(2017){Kopparapu}, {Wolf}, {Arney}, {Batalha},
  {Haqq-Misra}, {Grimm}, \& {Heng}}]{2017ApJ...845....5K}
{Kopparapu}, R.~k., {Wolf}, E.~T., {Arney}, G., {et~al.} 2017, \apj, 845, 5

\bibitem[{{Kreidberg} {et~al.}(2014){Kreidberg}, {Bean}, {D{\'e}sert},
  {Benneke}, {Deming}, {Stevenson}, {Seager}, {Berta-Thompson}, {Seifahrt}, \&
  {Homeier}}]{2014Natur.505...69K}
{Kreidberg}, L., {Bean}, J.~L., {D{\'e}sert}, J.-M., {et~al.} 2014, \nat, 505,
  69

\bibitem[{{Krissansen-Totton} {et~al.}(2018{\natexlab{a}}){Krissansen-Totton},
  {Garland}, {Irwin}, \& {Catling}}]{2018AJ....156..114K}
{Krissansen-Totton}, J., {Garland}, R., {Irwin}, P., \& {Catling}, D.~C.
  2018{\natexlab{a}}, \aj, 156, 114

\bibitem[{{Krissansen-Totton} {et~al.}(2018{\natexlab{b}}){Krissansen-Totton},
  {Olson}, \& {Catling}}]{2018SciA....4.5747K}
{Krissansen-Totton}, J., {Olson}, S., \& {Catling}, D.~C. 2018{\natexlab{b}},
  Science Advances, 4, eaao5747

\bibitem[{{Lin} {et~al.}(2021){Lin}, {MacDonald}, {Kaltenegger}, \&
  {Wilson}}]{2021MNRAS.505.3562L}
{Lin}, Z., {MacDonald}, R.~J., {Kaltenegger}, L., \& {Wilson}, D.~J. 2021,
  \mnras, 505, 3562

\bibitem[{{Line} {et~al.}(2016){Line}, {Stevenson}, {Bean}, {Desert},
  {Fortney}, {Kreidberg}, {Madhusudhan}, {Showman}, \&
  {Diamond-Lowe}}]{2016AJ....152..203L}
{Line}, M.~R., {Stevenson}, K.~B., {Bean}, J., {et~al.} 2016, \aj, 152, 203

\bibitem[{{Lustig-Yaeger} {et~al.}(2019){Lustig-Yaeger}, {Meadows}, \&
  {Lincowski}}]{2019AJ....158...27L}
{Lustig-Yaeger}, J., {Meadows}, V.~S., \& {Lincowski}, A.~P. 2019, \aj, 158, 27

\bibitem[{{Mikal-Evans} {et~al.}(2021){Mikal-Evans}, {Crossfield}, {Benneke},
  {Kreidberg}, {Moses}, {Morley}, {Thorngren}, {Molli{\`e}re},
  {Hardegree-Ullman}, {Brewer}, {Christiansen}, {Ciardi}, {Dragomir},
  {Dressing}, {Fortney}, {Gorjian}, {Greene}, {Hirsch}, {Howard}, {Howell},
  {Isaacson}, {Kosiarek}, {Krick}, {Livingston}, {Lothringer}, {Morales},
  {Petigura}, {Schlieder}, \& {Werner}}]{2021AJ....161...18M}
{Mikal-Evans}, T., {Crossfield}, I. J.~M., {Benneke}, B., {et~al.} 2021, \aj,
  161, 18

\bibitem[{{Molli{\`e}re} {et~al.}(2019){Molli{\`e}re}, {Wardenier}, {van
  Boekel}, {Henning}, {Molaverdikhani}, \& {Snellen}}]{2019A&A...627A..67M}
{Molli{\`e}re}, P., {Wardenier}, J.~P., {van Boekel}, R., {et~al.} 2019, \aap,
  627, A67

\bibitem[{{Morley} {et~al.}(2017){Morley}, {Kreidberg}, {Rustamkulov},
  {Robinson}, \& {Fortney}}]{2017ApJ...850..121M}
{Morley}, C.~V., {Kreidberg}, L., {Rustamkulov}, Z., {Robinson}, T., \&
  {Fortney}, J.~J. 2017, \apj, 850, 121

\bibitem[{{Pidhorodetska} {et~al.}(2020){Pidhorodetska}, {Fauchez},
  {Villanueva}, {Domagal-Goldman}, \& {Kopparapu}}]{2020ApJ...898L..33P}
{Pidhorodetska}, D., {Fauchez}, T.~J., {Villanueva}, G.~L., {Domagal-Goldman},
  S.~D., \& {Kopparapu}, R.~K. 2020, \apjl, 898, L33

\bibitem[{{Pont} {et~al.}(2006){Pont}, {Zucker}, \&
  {Queloz}}]{2006MNRAS.373..231P}
{Pont}, F., {Zucker}, S., \& {Queloz}, D. 2006, \mnras, 373, 231

\bibitem[{{Roberge} \& {Moustakas}(2018)}]{2018NatAs...2..605R}
{Roberge}, A., \& {Moustakas}, L.~A. 2018, Nature Astronomy, 2, 605

\bibitem[{{Robinson} \& {Catling}(2012)}]{2012ApJ...757..104R}
{Robinson}, T.~D., \& {Catling}, D.~C. 2012, \apj, 757, 104

\bibitem[{{Robinson} {et~al.}(2017){Robinson}, {Fortney}, \&
  {Hubbard}}]{2017ApJ...850..128R}
{Robinson}, T.~D., {Fortney}, J.~J., \& {Hubbard}, W.~B. 2017, \apj, 850, 128

\bibitem[{{Robinson} {et~al.}(2014){Robinson}, {Maltagliati}, {Marley}, \&
  {Fortney}}]{2014PNAS..111.9042R}
{Robinson}, T.~D., {Maltagliati}, L., {Marley}, M.~S., \& {Fortney}, J.~J.
  2014, Proceedings of the National Academy of Science, 111, 9042

\bibitem[{{Schwieterman} {et~al.}(2018){Schwieterman}, {Kiang}, {Parenteau},
  {Harman}, {DasSarma}, {Fisher}, {Arney}, {Hartnett}, {Reinhard}, {Olson},
  {Meadows}, {Cockell}, {Walker}, {Grenfell}, {Hegde}, {Rugheimer}, {Hu}, \&
  {Lyons}}]{2018AsBio..18..663S}
{Schwieterman}, E.~W., {Kiang}, N.~Y., {Parenteau}, M.~N., {et~al.} 2018,
  Astrobiology, 18, 663

\bibitem[{{Sergeev} {et~al.}(2020){Sergeev}, {Lambert}, {Mayne}, {Boutle},
  {Manners}, \& {Kohary}}]{2020ApJ...894...84S}
{Sergeev}, D.~E., {Lambert}, F.~H., {Mayne}, N.~J., {et~al.} 2020, \apj, 894,
  84

\bibitem[{{Sing} {et~al.}(2016){Sing}, {Fortney}, {Nikolov}, {Wakeford},
  {Kataria}, {Evans}, {Aigrain}, {Ballester}, {Burrows}, {Deming},
  {D{\'e}sert}, {Gibson}, {Henry}, {Huitson}, {Knutson}, {Etangs}, {Pont},
  {Showman}, {Vidal-Madjar}, {Williamson}, \& {Wilson}}]{2016Natur.529...59S}
{Sing}, D.~K., {Fortney}, J.~J., {Nikolov}, N., {et~al.} 2016, \nat, 529, 59

\bibitem[{{Snellen} {et~al.}(2010){Snellen}, {de Kok}, {de Mooij}, \&
  {Albrecht}}]{2010Natur.465.1049S}
{Snellen}, I.~A.~G., {de Kok}, R.~J., {de Mooij}, E.~J.~W., \& {Albrecht}, S.
  2010, \nat, 465, 1049

\bibitem[{{Suissa} {et~al.}(2020){Suissa}, {Mandell}, {Wolf}, {Villanueva},
  {Fauchez}, \& {Kopparapu}}]{2020ApJ...891...58S}
{Suissa}, G., {Mandell}, A.~M., {Wolf}, E.~T., {et~al.} 2020, \apj, 891, 58

\bibitem[{{Tremblay} {et~al.}(2020){Tremblay}, {Line}, {Stevenson}, {Kataria},
  {Zellem}, {Fortney}, \& {Morley}}]{2020AJ....159..117T}
{Tremblay}, L., {Line}, M.~R., {Stevenson}, K., {et~al.} 2020, \aj, 159, 117

\bibitem[{{Turbet} {et~al.}(2018){Turbet}, {Bolmont}, {Leconte}, {Forget},
  {Selsis}, {Tobie}, {Caldas}, {Naar}, \& {Gillon}}]{2018A&A...612A..86T}
{Turbet}, M., {Bolmont}, E., {Leconte}, J., {et~al.} 2018, \aap, 612, A86

\bibitem[{{Wakeford} {et~al.}(2019){Wakeford}, {Lewis}, {Fowler}, {Bruno},
  {Wilson}, {Moran}, {Valenti}, {Batalha}, {Filippazzo}, {Bourrier},
  {H{\"o}rst}, {Lederer}, \& {de Wit}}]{2019AJ....157...11W}
{Wakeford}, H.~R., {Lewis}, N.~K., {Fowler}, J., {et~al.} 2019, \aj, 157, 11

\bibitem[{{Wang} {et~al.}(2018){Wang}, {Mawet}, {Hu}, {Ruane}, {Delorme}, \&
  {Klimovich}}]{2018JATIS...4c5001W}
{Wang}, J., {Mawet}, D., {Hu}, R., {et~al.} 2018, Journal of Astronomical
  Telescopes, Instruments, and Systems, 4, 035001

\bibitem[{{Wolf}(2017)}]{2017ApJ...839L...1W}
{Wolf}, E.~T. 2017, \apjl, 839, L1

\bibitem[{{Wunderlich} {et~al.}(2019){Wunderlich}, {Godolt}, {Grenfell},
  {St{\"a}dt}, {Smith}, {Gebauer}, {Schreier}, {Hedelt}, \&
  {Rauer}}]{2019A&A...624A..49W}
{Wunderlich}, F., {Godolt}, M., {Grenfell}, J.~L., {et~al.} 2019, \aap, 624,
  A49

\bibitem[{{Zhang} {et~al.}(2018){Zhang}, {Zhou}, {Rackham}, \&
  {Apai}}]{2018AJ....156..178Z}
{Zhang}, Z., {Zhou}, Y., {Rackham}, B.~V., \& {Apai}, D. 2018, \aj, 156, 178

\bibitem[{{Zhou} {et~al.}(2017){Zhou}, {Apai}, {Lew}, \&
  {Schneider}}]{2017AJ....153..243Z}
{Zhou}, Y., {Apai}, D., {Lew}, B.~W.~P., \& {Schneider}, G. 2017, \aj, 153, 243

\end{thebibliography}

\appendix

\section{Posterior distributions} \label{app:posteriors}

\begin{figure*}
\centering  
\includegraphics[width=\linewidth]{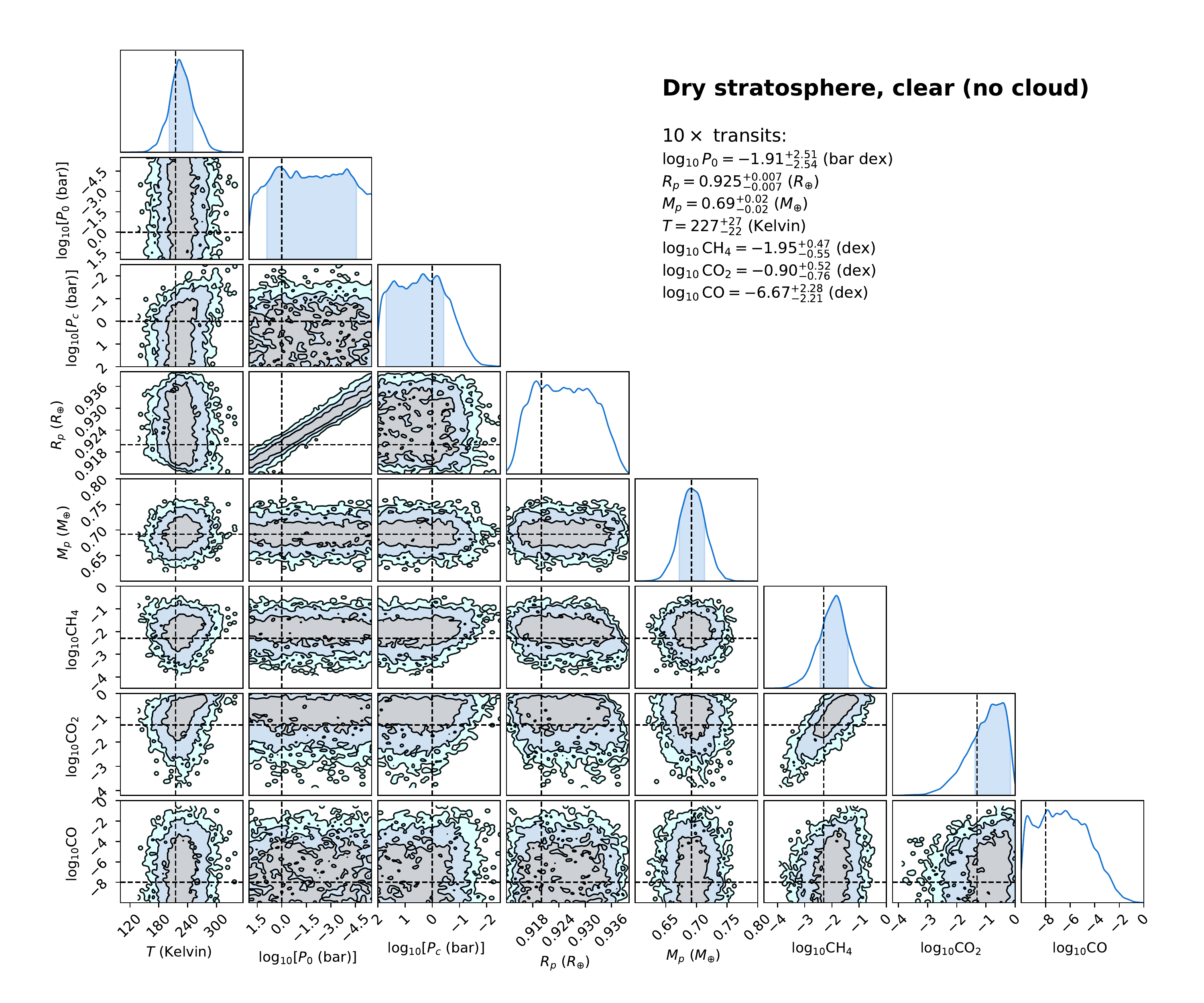}
\caption{Posterior distributions obtained for a subset of model parameters from the dry stratosphere retrieval analysis assuming a clear atmosphere and $N=10$ co-added transits. Panels along the diagonal show marginalised posterior distributions with blue shading indicating the $1\sigma$ credible ranges. Off-diagonal panels show parameter covariances with contours corresponding to the $1\sigma$, $2\sigma$, and $3\sigma$ credible bounds. Horizontal and vertical dashed black lines show the values for each parameter that were used to generate the synthetic datasets prior to performing the retrieval analyses. Text in the upper right half of the figure reports the posterior distribution medians and $1\sigma$ credible bounds.}
\label{fig:posterior2Dclear}
\end{figure*}

\begin{figure*}
\centering  
\includegraphics[width=\linewidth]{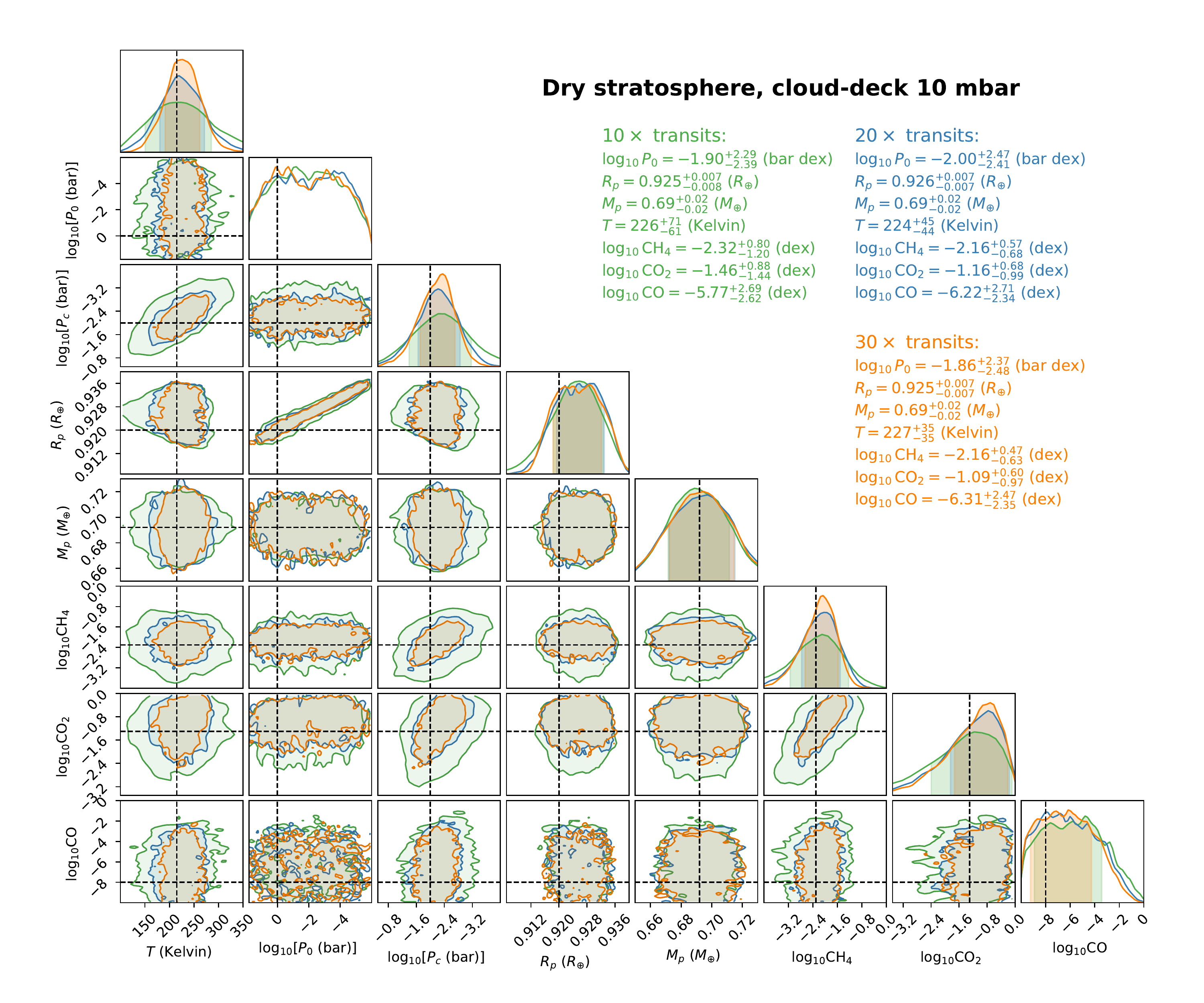}
\caption{Similar to Figure \ref{fig:posterior2Dclear}, but showing posterior distributions obtained for a subset of model parameters for the dry stratosphere scenario with a cloud/haze-layer at a pressure of 10\,mbar. Contours show the $1\sigma$ credible ranges obtained assuming $N=10$ (green), $20$ (blue), and $30$ (orange) co-added transits. Text in the upper right half of the figure reports the posterior medians and $1\sigma$ credible bounds, with the same colour-coding as the plotted posterior distributions.}
\label{fig:posterior2Dhz10mb}
\end{figure*}

For a subset of the most relevant parameters, Figure \ref{fig:posterior2Dclear} shows the posterior covariances and marginalised distributions that were obtained for the dry stratosphere scenario described in Section \ref{sec:datasets} assuming a clear atmosphere and $N=10$ co-added transits. Figure \ref{fig:posterior2Dhz10mb} shows the same, but for the dry stratosphere scenario with a cloud/haze-layer at a pressure of 10\,mbar and $N=10$, $N=20$, and $N=30$ co-added transits. These figures were produced using the publicly available \texttt{chainconsumer} Python software \citep{2016JOSS....1...45H}.

\end{document}